\documentclass[useAMS,usenatbib,fleqn]{mn2e}

\usepackage{amssymb}
\usepackage{times}
\usepackage{graphicx}
\usepackage{epstopdf}
\usepackage{subfigure}
\usepackage[T1]{fontenc}
\usepackage{aecompl} 
\usepackage{multirow}
\usepackage{pdflscape}
\usepackage{rotating}

\usepackage[fleqn]{amsmath}
\usepackage{mathrsfs}
\usepackage{xcolor}

\usepackage{anyfontsize}

\newcommand{\beq}{\begin{equation}}
\newcommand{\eeq}{\end{equation}}

\def\gs{\mathrel{\lower0.6ex\hbox{$\buildrel {\textstyle >}\over{\scriptstyle \sim}$}}}
\def\ls{\mathrel{\lower0.6ex\hbox{$\buildrel {\textstyle <}\over{\scriptstyle \sim}$}}}
\newcommand{\simgt}{\lower.5ex\hbox{$\; \buildrel > \over \sim \;$}}
\newcommand{\simlt}{\lower.5ex\hbox{$\; \buildrel < \over \sim \;$}}

\usepackage{hyperref}

\newcommand{\aap}{A\&A}

\newcommand{\apj}{ApJ}

\newcommand{\mnras}{MNRAS}

\newcommand{\ssr}{Space Science Reviews}

\defcitealias{se+et15_comalit_I}{CoMaLit-I}
\defcitealias{ser+al15_comalit_II}{CoMaLit-II}
\defcitealias{ser15_comalit_III}{CoMaLit-III}
\defcitealias{se+et15_comalit_IV}{CoMaLit-IV}
\defcitealias{xxl_I_pie+al15}{XXL-I}
\defcitealias{xxl_II_pac+al15}{XXL-II}
\defcitealias{xxl_III_gil+al15}{XXL-III}
\defcitealias{xxl_IV_lie+al15}{XXL-IV}
\defcitealias{xxl_XIII_eck+al15}{XXL-XIII}

\begin{document}

\title[LIRA]{A Bayesian approach to linear regression in astronomy}
\author[
M. Sereno
]{
Mauro Sereno$^{1}$\thanks{E-mail: mauro.sereno@unibo.it (MS)}
\\
$^1$Dipartimento di Fisica e Astronomia, Universit\`a di Bologna, viale Berti Pichat 6/2, 40127 Bologna, Italia\\
$^2$INAF, Osservatorio Astronomico di Bologna, via Ranzani 1, 40127 Bologna, Italia\\
}

%\date{August 27, 2015}

\maketitle

\begin{abstract}
Linear regression is common in astronomical analyses. I discuss a Bayesian hierarchical modeling of data with heteroscedastic and possibly correlated measurement errors and intrinsic scatter. The method fully accounts for time evolution. The slope, the normalization, and the intrinsic scatter of the relation can evolve with the redshift. The intrinsic distribution of the independent variable is approximated using a mixture of Gaussian distributions whose means and standard deviations depend on time. The method can address scatter in the measured independent variable (a kind of Eddington bias), selection effects in the response variable (Malmquist bias), and departure from linearity in form of a knee. I tested the method with toy models and simulations and quantified the effect of biases and inefficient modeling. The \textsc{R}-package \textsc{LIRA} (LInear Regression in Astronomy) is made available to perform the regression.
\end{abstract}


\section{Introduction}

%\upi

Astronomy and statistics have an interwoven history \citep{fe+ba12}. Linear regression is one of the most frequently used statistical techniques in astronomical data analysis. There is an impressive variety of methods to estimate functional relationships between variables. 

Linear regression is kind of easy. We can draw a line which nicely interpolates a distribution of points on a paper by eye. Connecting dots and forming a regular pattern is a game for kids. Difficulties lie in refining the results and uncovering the quantity we are really looking for. As an example, the ordinary least square estimator is elegant and powerful. Still, results may be meaningless if we apply it out of its range of validity.

Most astronomical data analyses feature intrinsic scatter about the regression line. Measurement errors can affect both the independent and the dependent variables. Errors may  be heteroscedastic, i.e., they differ, and possibly correlated. The intrinsic distribution of the independent variables may be irregular or not uniform. The independent variable may be hidden and we could measure just a proxy of it. Selection effects can make the observed sample not representative of the population we want to study.

These aspects influence regression results and can make the use of some statistical estimators inappropriate. Many methods have been proposed to tackle these effects \citep[ and references therein]{ak+be96,kel07,iso+al90,hog+al10,fe+ba12}. Here, we are mostly interested in methods assuming that the dispersions, either the intrinsic scatter or the uncertainties in the measurement process, are Gaussian. Non-Gaussian multivariate datasets need generalized linear methods \citep{des+al15}.

Some statistical papers had the great merit to clarify the involved problematics to the astronomical community. \citet{ak+be96} proposed the BCES estimator (Bivariate Correlated Errors and intrinsic Scatter) which accommodates intrinsic scatter in addition to correlated, heteroscedastic measurement errors on both variables by correcting the observed moments of the data.

\citet{kel07} described a Bayesian method (MLINMIX) based on the likelihood function of the measured data. The method can account for measurement errors, intrinsic scatter, multiple independent variables, non-detections, and selection effects in the independent variable. \citet{kel07} emphasized that the underlying distribution of covariates in a regression has to be modeled to get unbiased regression parameters and he proposed to approximate the intrinsic distribution of the independent variables as a mixture of Gaussian functions. This modeling is flexible when estimating the distribution of the true values of the independent variable and it is robust against model mispecification. 

Recently, \citet{man15} extended the MLINMIX algorithm to the case of multiple response variables and he described how to model the prior distribution of covariates using a Dirichlet process rather than a mixture. Alternative approaches based on generative models for the data were proposed too \citep{hog+al10,ro+ob15}.

Here, we build upon these methods to develop a linear regression tool optimized to the study of time evolving scaling relations. Some remarkable features show up in astronomical studies. As discussed above, astronomical data sets can be affected by heteroscedatic, correlated errors in both variables and by intrinsic scatter around the regression line.

Furthermore, the scaling parameters of the studied phenomenological relationship may not be constant with time. The source of intrinsic scatter is the variation of the physical properties which can be time dependent. The slope of the relation may be time dependent too if some physical processes are more conspicuous either early or lately. Time has then a special role and cannot be treated as a simple independent variable in a multivariate analysis.

In astronomical analyses, we are often interested in the correlation among an observable quantity against a variable which we do not have access to, e.g., the mass of a black hole, the mass of a galaxy cluster, the star formation rate of a galaxy. We cannot really measure these quantities but just scattered proxies of them, e.g., the weak lensing mass of a galaxy in place of the true mass.

Differently from gravity, a lot of astrophysical phenomena are scale dependent. Some baryonic processes may be triggered above some thresholds and be ineffective below. This can break linearity.

Selection effects and heterogeneity can make the astronomical sample used in the regression not representative of the population we are interested in. The sample may be sparse or selected according to the value of the response variable, as in a flux limited survey.

I discuss a hierarchical Bayesian method to deal with the above aspects. Its main assumption is that scatters and uncertainties are Gaussian. Part of it has been already presented and employed in the CoMaLit (COmparing MAsses in Literature) series of papers (\citealt[CoMaLit-I]{se+et15_comalit_I}, \citealt[CoMaLit-II]{ser+al15_comalit_II}, \citealt[CoMaLit-IV]{se+et15_comalit_IV}).

The method shares important features with other recently developed methods. \citet{mau14} proposed a model to constrain simultaneously the form and evolution of the scaling relations. The method distinguish between measured values, intrinsic scattered values, and model values and can constrain the intrinsic scatter and its covariance. Correlation among intrinsic scatters has to be considered in multivariate analyses to obtain unbiased scaling relations \citep{evr+al14,roz+al14c,man+al10,man+al15}. 

The paper is as follows. Ins Sec.~\ref{sec_evol}, I introduce power-law scaling relations and their linear counterparts in logarithmic space. The hierarchical Bayesian model is presented in Sec.~\ref{sec_regr}. Section~\ref{sec_simu} is devoted to simulations and algorithm testing. Final considerations are in Sec.~\ref{sec_simu}. Appendix~\ref{app_pack} gives some information on the package accompanying the paper. Appendix~\ref{app_reds} gives some hints about the modeling of the time evolution.

If needed, I adopt the same conventions and notations of the CoMaLit series. The frame-work cosmological model is the concordance flat $\Lambda$CDM universe with matter density parameter $\Omega_\text{M}=0.3$; $H(z)$ is the redshift dependent Hubble parameter and $E_z\equiv H(z)/H_0$.  `$\log$' is the logarithm to base 10 and `$\ln$' is the natural logarithm.

The method described in the present paper has been implemented in the R language\footnote{\url{http://www.r-project.org}}. The package is named LIRA (LInear Regression in Astronomy) and it is publicly available from CRAN (Comprehensive R Archive Network) or GitHub, see App.~\ref{app_pack}.

\section{Linear scaling}
\label{sec_evol}

Most of the scaling relations we deal with in astronomy are time evolving power-laws. This simple schematism is supported by observations, theoretical considerations, and numerical simulations  \citep{sta+al10,gio+al13}. The general form of the relation between two properties, e.g., the observable $O$ and the mass $M$, is 
\beq
\label{eq_evo_1}
O \propto M^\beta F_z^\gamma,
\eeq
where  $\beta$ is the slope and the redshift evolution in the median scaling relation is accounted for by the factor $F_z$. According to the context, the redshift factor $F_z$ may be either $E_z$ or the factor $(1+z)$. In logarithmic variables, the scaling relation is linear and the scatter is Gaussian, 
\beq
\label{eq_evo_2}
\log O = \alpha+ \beta \log M + \gamma \log F_z.
\eeq
In the following, $T=\log F_z$. If spectroscopically determined, measurement uncertainties in redshift are negligible\footnote{I am not considering catastrophic errors.}. The relative uncertainty in photometric redshifts can be small too, and usually smaller than uncertainties in other measurable properties, such as mass, luminosity, or temperature. I will not consider redshift uncertainties in the following.

In the usual framework, the time evolution does not depend on the mass scale and only affects the normalization. This is supported by the self-similar scenario \citep{gio+al13}, where the factor $F_z$ for observable properties measured within the same over-density radius is $E_z$. However, the interplay between different physical processes that can be more or less effective at different times and can make the slope time dependent, $\beta (z)$. Assuming that the evolution of the slope with redshift is linear in $T$, Eq.~(\ref{eq_evo_2}) can be generalized as
\beq
\label{eq_evo_3}
Y = \alpha+ \beta \ X + \gamma \ T + \delta \ X \  T.
\eeq
where $X=\log M$, and $Y=\log O$. $X$ and $Y$ are random latent variables with intrinsic scatter. The time variable $T$ is deterministic, not affected by measurement errors (which I neglect). Latent and observed variables are related through a structural equation model \citep[ and references therein]{kel07}. In statisticians' terms, the critical criterion is linearity in the model parameters, not in the model variables, which makes  Eq.~(\ref{eq_evo_3}) a linear model.

\section{Regression scheme}
\label{sec_regr}

\begin{table*}
\caption{Parameters of the regression scheme and their description. The variables $Z$ is the covariate, $X$ is a proxy of $Z$, and $Y$ is the response variable. $z=z_\text{ref}$ is the user defined reference redshift. $D$ is either the luminosity or angular diameter distance. $F_z$ and $D$ are normalized such that $F_z(z_\text{ref})=1$ and $D(z_\text{ref})=1$. \textsc{LIRA} is highly customizable and priors can be set by the user among the distributions defined in JAGS. In addition, \textsc{LIRA} defines the \texttt{prec.dgamma} prior too. Priors in square brackets can be set only as delta distributions. See Sec.~\ref{sec_prio} and the \textsc{LIRA} user manual for details.}
\label{tab_par}
\centering
\resizebox{\hsize}{!} {
\begin{tabular}[c]{l l l l l}
	\hline
	Type & Meaning & Symbol & Code symbol & Default prior\\ 
	\noalign{\smallskip}  
	\hline
	\multicolumn{5}{c}{$Y$-$Z$ scaling} \\
	 \noalign{\smallskip}  
	\multicolumn{5}{l}{$Y_Z = \alpha_{Y|Z}+\beta_{Y|Z} Z + \gamma_{Y|Z} T+ \delta_{Y|Z} Z\ T$} \\
	\noalign{\smallskip}  
	Conditional scaling relation &	intercept & $\alpha_{Y|X}$ & \texttt{alpha.YIZ} & \texttt{dunif}\\
	 & slope & $\beta_{Y|X}$  & \texttt{beta.YIZ} & \texttt{dt}\\
	& time evolution & $\gamma_{Y|Z}$  & \texttt{gamma.YIZ} & \texttt{dt}\\
	& time tilt & $\delta_{Y|Z}$  & \texttt{delta.YIZ} & \texttt{0}\\
	\noalign{\smallskip}
	\multicolumn{5}{l}{$Y_Z = \alpha_{Y|Z,\text{knee}}+\beta_{Y|Z,\text{knee}} Z + \gamma_{Y|Z,\text{knee}} T+ \delta_{Y|Z,\text{knee}} Z\ T$} \\
	\noalign{\smallskip}  
	Scaling relation before the break  & slope for $Z<Z_\text{knee}$    & $\beta_{Y|X,\text{knee}}$  & \texttt{beta.YIZ.knee} & \texttt{beta.YIZ}\\
		                                          & time tilt  for $Z<Z_\text{knee}$    & $\delta_{Y|Z,\text{knee}}$  & \texttt{delta.YIZ.knee} & \texttt{delta.YIZ}\\
	\noalign{\smallskip}
	\multicolumn{5}{l}{$f_\text{knee}(Z)=1/(1+\exp [(Z-Z_\text{knee})/l_\text{knee} ])$} \\
	\noalign{\smallskip}  
	Transition function  & break scale    & $Z_\text{knee}$  & \texttt{Z.knee} & \texttt{dunif}\\
		                       & break length   & $l_\text{knee}$   & \texttt{l.knee} & \texttt{1e-04}\\
	\multicolumn{5}{c}{$X$-$Z$ scaling} \\
	\noalign{\smallskip}
		\multicolumn{5}{l}{$X_Z = \alpha_{X|Z}+\beta_{X|Z} Z + \gamma_{X|Z} T + \delta_{X|Z} Z\ T$} \\
	\noalign{\smallskip}  
	Proxy of the independent variable &	bias & $\alpha_{X|Z}$ & \texttt{alpha.XIZ} & \texttt{0}\\
	 & slope & $\beta_{X|Z}$  & \texttt{beta.XIZ} & \texttt{1}\\
	& time evolution & $\gamma_{X|Z}$  & \texttt{gamma.XIZ} & \texttt{0}\\
	& time tilt & $\delta_{X|Z}$  & \texttt{delta.XIZ} & \texttt{0}\\
	\noalign{\smallskip}
	\hline
	\multicolumn{5}{c}{Scatters} \\
	\noalign{\smallskip}
	\multicolumn{5}{l}{$\sigma_{Y|Z}=[\sigma_{Y|Z,0} + f_\text{knee}(Z)(\sigma_{Y|Z,0,\text{knee}}-\sigma_{Y|Z,0}) ]F_z^{\gamma_{\sigma_{Y|Z,F_z}}}D^{\gamma_{\sigma_{Y|Z,D}}}$} \\  
	 \noalign{\smallskip}  
	 Intrinsic scatter & scatter at $z=z_\text{ref}$ for $Z\ge Z_\text{knee}$	& $\sigma_{Y|Z,0}$ & \texttt{sigma.YIZ.0} & \texttt{prec.dgamma}\\
	                                            & scatter at $z=z_\text{ref}$ for $Z<Z_\text{knee}$	& $\sigma_{Y|Z,0,\text{knee}}$ & \texttt{sigma.YIZ0.knee} & \texttt{sigma.YIZ.0}\\
	 &	time evolution	with $F_z$ 			& $\gamma_{\sigma_{Y|Z,F_z}}$ & \texttt{gamma.sigma.YIZ.Fz} & \texttt{0} \\
	&	time evolution	with $D$	         	         & $\gamma_{\sigma_{Y|Z,D}}$ & \texttt{gamma.sigma.YIZ.D} & \texttt{0} \\
	\noalign{\smallskip}  
	\multicolumn{5}{l}{$\sigma_{X|Z}=\sigma_{X|Z,0} F_z^{\gamma_{\sigma_{X|Z,F_z}}}D^{\gamma_{\sigma_{X|Z,D}}}$} \\  
	 \noalign{\smallskip}  
	 Intrinsic scatter of the proxy  & scatter at $z=z_\text{ref}$	& $\sigma_{X|Z,0}$ & \texttt{sigma.XIZ.0} & \texttt{0}\\
	 & time evolution with $F_z$				& $\gamma_{\sigma_{X|Z,F_z}}$ & \texttt{gamma.sigma.XIZ.Fz} & \texttt{0}\\
	  & time evolution with $D$		& $\gamma_{\sigma_{X|Z,F_z}}$ & \texttt{gamma.sigma.XIZ.D} & \texttt{0 }\\
	\noalign{\smallskip}
	\multicolumn{5}{l}{$\rho_{XY|Z}  =  \rho_{XY|Z,0}F_z^{\gamma_{\rho_{XY|Z,F_z}}}D^{\gamma_{\rho_{XY|Z,D}}}$} \\  
	 \noalign{\smallskip}  
	 Intrinsic scatter correlation  & correlation at $z=z_\text{ref}$	& $ \rho_{XY|Z,0}$ & \texttt{rho.XYIZ.0} & \texttt{0}\\
	 & time evolution with $F_z$				& $\gamma_{\rho_{XY|Z,F_z}}$ & \texttt{gamma.rho.XYIZ.Fz} & \texttt{0}\\
	 & time evolution with $D$		& $\gamma_{\rho_{XY|Z,F_z}}$ & \texttt{gamma.rho.XYIZ.D} & \texttt{0 }\\
	 \hline  
	 \noalign{\smallskip}
	 \multicolumn{5}{c}{Intrinsic  distribution of the independent variable}  \\
	 \noalign{\smallskip}  
	\multicolumn{5}{l}{$p(Z) \propto \left[  \sum_k \pi_k \ {\cal N}(\mu_{Z,k} (z), \sigma_{Z,k}(z)) \right]  {\cal U}(Z_\text{min},Z_\text{max})$}  \\
	 \noalign{\smallskip}  
	  Gaussian mixture & number of components   & $n_\text{mix}$ & \texttt{n.mixture} & \texttt{[1]}\\
	   & weights of the components   & $\pi_{k}$ & \texttt{pi[k]} & \texttt{ddirch }\\
	   & minimum  $Z$ value (only for $n_\text{mix}=1$) & $Z_\text{min}$ & \texttt{Z.min} & \texttt{$-\infty$ }\\
	   & maximum  $Z$ value  (only for $n_\text{mix}=1$)& $Z_\text{max}$ & \texttt{Z.max} & \texttt{$+\infty$ }\\
	\multicolumn{5}{l}{$\mu_{Z,k} (z) = \mu_{Z,0k} +\gamma_{\mu_Z,F_z}T + \gamma_{\mu_Z,D}\log D$} \\ 
	 \noalign{\smallskip}  
	  Means of the Gaussian & mean of the first component at $z=z_\text{ref}$ & $\mu_{Z,01}$ & \texttt{mu.Z.0} & \texttt{dunif }\\
	  components                       & means of the additional components & $\mu_{Z,0k}$ & \texttt{mu.Z.0.mixture[k]}  & \texttt{dunif }\\
	    &  ($2\le k \le n_\text{mix}$)  at $z=z_\text{ref}$ &  &  & \\
	  & time evolution with $F_z$ & $\gamma_{\mu_Z,F_z}$ & \texttt{gamma.mu.Z.Fz} & \texttt{dt } \\
	 & time evolution with $D$    & $\gamma_{\mu_Z,D}$     & \texttt{gamma.mu.Z.D} & \texttt{dt } \\
	 \noalign{\smallskip}  
	\multicolumn{5}{l}{ $\sigma_{Z,k}(z)=\sigma_{Z,0k}F_z^{\gamma_{\sigma_{Z},F_z}}D^{\gamma_{\sigma_{Z},D}}$} \\
        \noalign{\smallskip}  
       Standard deviations of the & deviation of the first component at $z=z_\text{ref}$	& $\sigma_{Z,01}$ & \texttt{sigma.Z.0} & \texttt{prec.dgamma }\\
       Gaussian  components & deviations of the additional components 	& $\sigma_{Z,0k}$ & \texttt{sigma.Z.0.mixture[k]} & \texttt{prec.dgamma }\\
               &  ($2\le k \le n_\text{mix}$)  at $z=z_\text{ref}$ &  &  & \\
       & time evolution with $F_z$	& $\gamma_{\sigma_{Z},F_z}$ & \texttt{gamma.sigma.Z.Fz} & \texttt{0}\\
       & time evolution with $D$	& $\gamma_{\sigma_{Z},D}$ & \texttt{gamma.sigma.Z.D} & \texttt{0}\\
	\hline
	\end{tabular}
	}
\end{table*}

The Bayesian regression model presented in the following is a measurement error model \citep{fe+ba12}. Measurement errors are involved in the hierarchical structure and incorporated into the model. I assume that all scatter terms, i.e., intrinsic scatter and measurement errors, are Gaussian with zero mean although with different variances. 

Linear regression in astronomy is usually characterized by intrinsic scatter around the scaling relation and measurement errors in both the independent and the dependent variables. I assume that the covariate variable $X_Z$ and the response variable $Y_Z$, which are latent, fall exactly on a straight line. This is the underlying relation we want to discover. The latent variables cannot be measured. We can measure their proxies $X$ and $Y$, which differ from $X_Z$ and $Y_Z$ for the intrinsic scatters. These are intrinsic deviations of data points from the intrinsic scaling relation that are present even if all measurements were made with perfect precision and accuracy. 

The proxies $X$ and $Y$ are linked to the observed manifest variables $x$ and $y$ with additional error terms. We could measure $X$ and $Y$ only in an ideal experiment with infinite accuracy and precision.

The measured values of $x$ and $y$ and their known measurement errors are the inputs to the model. The variables $X$, $Y$, $X_Z$ and $Y_Z$ have to be determined in the regression procedure. The regression scheme is summarized in Table~\ref{tab_par} and described in details in the following.

\subsection{Linear scaling}

The linear relation between two unscattered quantities (I am not counting the time) can be expressed as
\beq
Y_Z  =  \alpha_{Y|Z}+\beta_{Y|Z} Z + \gamma_{Y|Z} T + \delta_{Y|Z} Z\ T, \label{eq_bug_1}
\eeq
where $\alpha$ denotes the normalization, the slope $\beta$ accounts for the dependence with $Z$, the slope $\gamma$ accounts for the time-evolution of the normalization and $\delta$ quantifies the tilt of the slope with time.

The basic case summarized in Eq.~(\ref{eq_bug_1}) is enough to describe the regression of $Y_Z$ against a variable which is directly observable. This is the case of the luminosity versus temperature relation of galaxy clusters. In some other cases, the independent variable $Z$ is not directly available from observations. For example, we cannot measure the mass of a cluster ($Z$), but we can approximate it with the weak lensing mass ($X$). We have then to couple Eq.~(\ref{eq_bug_1}) with
\beq
X_Z  =  \alpha_{X|Z}+\beta_{X|Z} Z + \gamma_{X|Z} T + \delta_{X|Z} Z\ T, \label{eq_bug_2}
\eeq
In this case, $X_Z$ and $Y_Z$ are related to the same covariate variable, $Z$.  The relations among $Z$, $X_Z$ and $Y_Z$ are deterministic and they are not affected by scatter. $X_Z$ and $Y_Z$ are rescaled versions of the latent variable $Z$, which can be seen as a fundamental property of the object, e.g., the mass of a cluster of galaxies.

\subsection{Intrinsic scatter}

The intrinsic scatter quantifies how close the data distribution is to strict linearity. The true properties of an astronomical object $X$ and $Y$, which we can try to measure, are intrinsically scattered with respect to the latent model variables $X_Z$ and $Y_Z$, which fall on a line without deviations but which are hidden properties. 

Observable properties are usually log-normally distributed about the mean scaling relations \citep{sta+al10,fab+al11,ang+al12,sar+al13}. This is supported by numerical simulations \citep{sta+al10,fab+al11,ang+al12} and observational studies \citep{mau07,vik+al09}. We assume that the intrinsic scatters are Gaussian,
\beq
P(\mathbfit{X}_i,\mathbfit{Y}_i | \mathbfit{X}_{Z_i},\mathbfit{Y}_{Z,i} ) = {\cal N}^\text{2D}(\{\mathbfit{X}_{Z,i},\mathbfit{Y}_{Z,i}\},\mathbfss{V}_{\sigma,i}), \label{eq_bug_4}
\eeq
where ${\cal N}^\text{2D}$ is the bivariate Gaussian distribution and $\mathbfss{V}_{\sigma,i}$ is the scatter covariance matrix of the $i$-th cluster whose diagonal elements are denoted as $\sigma_{X|Z,i}^2$ and $\sigma_{Y|Z,i}^2$, and whose off-diagonal elements are denoted as $\rho_{XY|Z,i}\sigma_{X|Z,i}\sigma_{Y|Z,i}$.

The intrinsic scatter of a scaling relation is related to the degree of regularity of the sample. The scatter can be prominent in morphologically complex halos or in objects which depart from dynamical/hydrostatic equilibrium \citep{fab+al11,sar+al13}. Deviations from spherically symmetry are another major source of scatter \citep{lim+al13,ser+al13}. Since high redshift objects are more irregular and less spherical, the scatter is usually expected to increase with redshift \citet{sar+al13, fab+al11}. The degree of scatter and  its evolution depends on the baryonic physics too \citep{fab+al11}. 

The time evolution of the scatters and of their correlation can be modeled as \citepalias{se+et15_comalit_IV}
\begin{align}
\sigma_{X|Z}(z) & = \sigma_{X|Z,0}F_z^{\gamma_{\sigma_{X|Z,F_z}}}D^{\gamma_{\sigma_{X|Z,D}}}, \label{eq_bug_5} \\
\sigma_{Y|Z}(z) & =  \sigma_{Y|Z,0}F_z^{\gamma_{\sigma_{Y|Z,F_z}}}D^{\gamma_{\sigma_{Y|Z,D}}}, \label{eq_bug_6} \\
\rho_{XY|Z}(z) & =  \rho_{XY|Z,0}F_z^{\gamma_{\rho_{XY|Z,F_z}}}D^{\gamma_{\rho_{XY|Z,D}}},  \label{eq_bug_7}
\end{align}
where $D$ is either the luminosity or the angular diameter distance.

If we want to regress $Y$ against $X$, we can identify $X$ and $Z$. Equation~(\ref{eq_bug_4}) reduces to
\beq
P(\mathbfit{Y}_i | \mathbfit{Z}_i  ) = {\cal N} ( \mathbfit{Y}_{Z,i} ,\sigma_{Y|Z,i}^2), \label{eq_bug_4b}
\eeq

\subsection{Measurement uncertainties}

The measured quantities \mathbfit{x} and \mathbfit{y} are the manifest values of  \mathbfit{X} and \mathbfit{Y}\footnote{\mathbfit{x}, \mathbfit{y}, \mathbfit{X}, and \mathbfit{Y} are vectors of $n$ elements.}. Due to observational uncertainties their relation can be expressed as
\beq
P(\mathbfit{x}_i,\mathbfit{y}_i | \mathbfit{X}_i,\mathbfit{Y}_i) = {\cal N}^\text{2D}(\{\mathbfit{X}_i,\mathbfit{Y}_i\},\mathbfss{V}_{\delta,i}), \label{eq_bug_3}
\eeq
where $\mathbfss{V}_{\delta,i}$ is the uncertainty covariance matrix whose diagonal elements are denoted as $\delta_{x,i}^2$ and $\delta_{y,i}^2$, and whose off-diagonal elements are denoted as $\rho_{xy,i}\delta_{x,i}\delta_{y,i}$. 
%We denote the intrinsic scatter as $\sigma$ and the measurement uncertainty as $\delta$. 

As a result of the $i$-th measurement process, we obtain $\mathbfit{x}_i$, $\mathbfit{y}_i$ and the related uncertainty covariance matrix  $\mathbfss{V}_{\delta,i}$. The variables $\mathbfit{X}_{Z,i}$, $\mathbfit{Y}_{Z,i}$, $\mathbfit{X}_i$, and $\mathbfit{Y}_i$, are unknown variables to be determined under the assumption of linearity.

\subsection{Malmquist bias}

Selection effects are a common concern in the astronomical analysis. If only objects above an observational threshold (in the response variable) are included, the sample is affected by the Malmquist bias \citep{mal20}. In this case, the relation between the measured and the true values (Eq.~\ref{eq_bug_3}) or between the true values and the unscattered values (Eq.~\ref{eq_bug_4}) has to be modified. 

The bias can be modeled by truncating the probability distributions below the threshold $\mathbfit{y}_\text{th,i}$. The measured and the true values of the quantities are now related as
\beq
\label{eq_MB_1}
P(\mathbfit{x}_i, \mathbfit{y}_i | \mathbfit{X}_i, \mathbfit{Y}_i) \propto {\cal N}^\text{2D}(\{ \mathbfit{X}_i, \mathbfit{Y}_i\},\mathbfss{V}_{\delta,i}) {\cal U}(\mathbfit{y}_\text{th,i},), 
\eeq
where ${\cal U}$ is the uniform distribution null for $y<\mathbfit{y}_\text{th,i}$. 

The observational thresholds $ \mathbfit{y}_\text{th}$ may not be exactly known. This may be the case when the quantity which the selection procedure is based on differs from the quantity used in the regression. We have then to consider the additional relation
\beq
 \label{eq_MB_2}
P(\mathbfit{y}_\text{th,i})={\cal N}(\mathbfit{y}_\text{th,obs,i},\delta_{y_\text{th,i}}^2),
\eeq
where $\delta_{y_\text{th,i}}$ is the uncertainty associated to the measured threshold $ \mathbfit{y}_\text{th,obs,i}$. Equations~(\ref{eq_MB_1} and \ref{eq_MB_2}) can be combined by considering a sigmoid curve instead of the step function in Eq.~(\ref{eq_MB_1}). 

The conditional probability of the proxies in the sample is truncated too,
\beq
\label{eq_MB_3}
P(\mathbfit{X}_i, \mathbfit{Y}_i | \mathbfit{X}_{Z,i}, \mathbfit{Y}_{Z,i} ) \propto {\cal N}^\text{2D}(\{ \mathbfit{X}_{Z,i}, \mathbfit{Y}_{Z,i}\},\mathbfss{V}_{\sigma,i}){\cal U}(\mathbfit{Y}_\text{th,i},), 
\eeq
where the threshold $\mathbfit{Y}_\text{th,i}$ follows the distribution
\beq
\label{eq_MB_4}
P(\mathbfit{Y}_\text{th,i})={\cal N}(\mathbfit{y}_\text{th,i},\delta_{y_i}^2) .
\eeq
If the thresholds are known without uncertainties, we can identify $\mathbfit{y}_\text{th}$ and $\mathbfit{Y}_\text{th}$ and the right-hand side of Eq.~(\ref{eq_MB_4}) is formally substituted by a Dirac delta function.

\subsection{Intrinsic distribution of the covariate}

The proper modeling of the distribution of the independent variable is crucial \citep{kel07}. Samples considered in regression analyses are usually biased with respect to the parent population. Sources may be selected according to their properties. Furthermore, even in absence of selection effect the intrinsic parent population is usually not uniform, which may cause border effects.

The intrinsic distribution of the independent variable $Z$  is shaped by two main effects. On one hand, very massive objects are rarer. On the other hand, massive objects are usually strong emitters and are easier to be detected to very large distances. As a result, the shape of the distribution is fairly unimodal and it evolves with time \citepalias{se+et15_comalit_IV}. 

The combined evolution of the completeness  and of the parent population can be characterized through the evolution of the peak and of the dispersion of the distribution of the selected sample. The intrinsic distribution of $Z$ can be approximated with a mixture of $n_\text{mix}$ time-evolving Gaussian functions  (\citealt{kel07}, \citetalias{ser+al15_comalit_II}, \citetalias{se+et15_comalit_IV}),
\beq
p(Z) = \sum_{k=1}^{n_\text{mix}} \pi_k \ {\cal N} \left (\mu_{Z,k} (z), \sigma_{Z,k}^2(z) \right), \label{eq_mix_1}
\eeq
where $\pi_k$ is the probability of drawing a data point from the $k$-th component, $\sum_k \pi_k =1$. 

I assume that the mixture components have different mean and dispersion but share the same evolution parameters. The mean of each component is connected to the (redshift-evolving) observational thresholds and to the intrinsic scatter of the observable quantity used to select the clusters, which evolves too. As a result, the evolution of the (mean of the) $k$-th mixture component can be modeled as \citepalias{se+et15_comalit_IV},
\beq
\label{eq_mix_2}
\mu_{Z,k} (z) = \mu_{Z,0k} +\gamma_{\mu_Z,F_z}T + \gamma_{\mu_Z,D}\log D,
\eeq
where $\mu_{Z,0k}$ is the mean at the reference redshift.

The evolution of the dispersions is related to the intrinsic scatter of the observable property used to select the sample. The time dependence can be modeled as \citepalias{se+et15_comalit_IV}
\beq
\label{eq_mix_3}
\sigma_{Z,k}(z)=\sigma_{Z,0k}F_z^{\gamma_{\sigma_{Z},F_z}}D^{\gamma_{\sigma_{Z},D}}.
\eeq

The proper modeling of the intrinsic distribution of the independent variable is crucial to correct for the Eddington bias, when the average value of an observed sample differs from the true intrinsic average of the objects of the same class (\citealt{edd13,jef38,edd40};\citetalias{se+et15_comalit_I}).
%The Eddington bias was originally formulated to correct count statistics for the effect of measurement uncertainty \citep{edd13,jef38,edd40} but it can be effective also when the independent variable is affected by intrinsic scatter \citepalias{se+et15_comalit_I}.

The intrinsic distribution can be alternatively modeled as a truncated Gaussian distribution
\beq
p(Z) =  {\cal N} \left (\mu_{Z,0} (z), \sigma_{Z,0}^2(z) \right)\ {\cal U} (Z_\text{min},Z_\text{max}). \label{eq_mix_4}
\eeq

\subsection{Departure from linearity}

Physical processes are effective at different scales, which may cause deviation from linearity. Detection of changes in the characteristics of random processes is related to the sphere of statistical analysis called the theory of change-point detection \citep{br+da93,ki+ec14}.

Gravity is the driving force behind formation and evolution of galaxy clusters but at small scales baryonic physics can play a prominent role. As a result, linearity can break. This can be shaped with a knee in the relation, such that before the breaking scale $Z_\text{knee}$, the scaling follows
\beq
Y_Z  =  \alpha_{Y|Z,\text{knee}}+\beta_{Y|Z,\text{knee}} Z + \gamma_{Y|Z,\text{knee}} T +\delta_{Y|Z,\text{knee}} Z\ T . \label{eq_kne_1}
\eeq
The normalization $\alpha_{Y|Z,\text{knee}}$ and the time evolution $\gamma_{Y|Z,\text{knee}}$ are determined by requiring equality at the transition  $Z_\text{knee}$,
\begin{align}
\alpha_{Y|Z,\text{knee}}  & = \alpha_{Y|Z}+ (\beta_{Y|Z} -\beta_{Y|Z,\text{knee}}) \label{eq_kne_2} \\
\gamma_{Y|Z,\text{knee}} & = \gamma_{Y|Z} \label{eq_kne_3}.
\end{align}

The transition between the two regimes can be modeled through a transition function,
\beq
\label{eq_kne_4}
f_\text{knee}=\frac{1}{1+\exp\left[ (Z-Z_\text{knee})/l_\text{knee}\right]},
\eeq
where the scale $l_\text{knee}$ sets the transition length. The relation over the full range reads
\begin{multline}
\label{eq_kne_5}
Y_Z = \alpha_{Y|Z}+\beta_{Y|Z}Z + \gamma_{Y|Z} T + \delta_{Y|Z}Z\ T  + (Z_\text{knee}-Z) f_\text{knee}(Z) \\
 \times \left[ (\beta_{Y|Z} -\beta_{Y|Z,\text{knee}})+ (\delta_{Y|Z} - \delta_{Y|Z,\text{knee}})\ T  \right] ,
\end{multline}
The same physical processes can affect the scatter too, which I model as
\beq
\label{eq_kne_6}
\sigma_{Y|Z}(Z, z_\text{ref})=\sigma_{Y|Z,0}+ (\sigma_{Y|Z,0,\text{knee}}-\sigma_{Y|Z,0}) f_\text{knee}(Z) .
\eeq
I assume that the redshift evolution of the scatter is not affected.

\subsection{Priors}
\label{sec_prio}

The Bayesian statistical treatment requires the explicit declaration of the priors. Priors can be either conveniently non-informative, if we have no guess on the parameters \citepalias{se+et15_comalit_I,ser+al15_comalit_II}, or peaked and with small dispersion, to convey the information obtained with concluded experiments or theory. The parameters can be also frozen by fixing them with a Dirac delta-prior. Priors in \textsc{LIRA} are highly customizable. In the following, I list the default choices.

Standard priors on the intercept $\alpha_{Y|Z}$ and on the means of the mixture components $\mu_{Z,0,k}$ can be flat,
\beq
\label{eq_pri_1}
\alpha_{Y|Z},\  \mu_{Z,0,k}  \sim  {\cal U}(-n_\text{L},n_\text{L}),
\eeq
where $n_\text{L}$ is a large number\footnote{In \textsc{LIRA}, $n_\text{L}$ is customizable and the shortcut for the prior is \texttt{dunif}. The default value is $n_\text{L} = 10^{4}$.}.

The slopes can follow the Student's $t_1$ distribution with one degree of freedom, as suitable for uniformly distributed direction angles\footnote{In \textsc{LIRA}, the shortcut for this prior is \texttt{dt}.},
\beq
\label{eq_pri_2}
\beta_{Y|Z},\ \gamma_{Y|Z},\ \gamma_{\mu_Z,F_z},\ \gamma_{\mu_Z,D} \sim  t_1.
\eeq
The $\gamma$-type parameters are set to zero when no redshift information is provided. The other slope parameters ($\beta_{X|Z}$, the $\delta$'s, the other $\gamma$'s) are by default frozen to 0. They can be unpegged by setting other priors. In these cases, the non-informative $t_1$ prior is suggested.

For the variance, I adopted by default a scaled inverse $\chi^2$-distribution\footnote{The default inverse $\chi^2$ prior for the variance is equivalent to a Gamma distribution $\Gamma(r=1/n_\text{L},\lambda=1/n_\text{L})$ for the precision, i.e., the inverse of the variance. Hence, the name \texttt{prec.dgamma} for this prior in \textsc{LIRA}, where \texttt{prec.dgamma} is applicable only to variances. Other customizable priors model directly the standard deviation.},
\beq
\label{eq_pri_3}
\sigma_{Y|Z,0}^2,\ \sigma_{Z,0,k}^2 \sim \text{Scale-inv-}\chi^2(\nu,\xi),
\eeq
with $\nu=2/n_\text{L}$ degrees of freedom and scale $\xi=1$. This parameter choice makes the distribution in Eq.~(\ref{eq_pri_3}) nearly scale-invariant.

%For the precision, i.e. the inverse of the variance, I adopted by default a nearly scale-invariant Gamma distribution,
%\beq
%\label{eq_pri_3b}
%1/\sigma_{Y|Z,0}^2,\ 1/\sigma_{Z,0,k}^2 \sim \Gamma(r,\lambda),
%\eeq
%where the rate $r$ and the shape parameter $\lambda$ of the $\Gamma$ distribution are fixed as $r=\lambda=1/n_\text{L}$. In \textsc{LIRA}, the shortcut for this prior is \texttt{prec.dgamma}. 

By default, $X$ tallies with $Z$ and it is unscattered, $\sigma_{X|Z,0}=0$. The scatter correlation  $\rho_{XY|Z,0}$ is set to zero too. Otherwise, a flat prior can be adopted,
\beq
\label{eq_pri_4}
\rho_{XY|Z,0} \sim  {\cal U}(-1,1).
\eeq

The parameters in the scaling $Y$-$Z$ and $X$-$Z$, see Eqs.~(\ref{eq_bug_1} and \ref{eq_bug_2}) are redundant. If we do not know the value of $Z$, we cannot measure all of them. By default, \textsc{LIRA} assumes that $X$ is an unbiased proxy of $Z$, i.e.,  $\alpha_{X|Z}=0$, $\beta_{X|Z}=1$, $\gamma_{X|Z}=0$, and $\delta_{X|Z}=0$. For linear relations, fixing the parameters of the $X$-$Z$ rather than the $Y$-$Z$ relation is just a matter of rescaling. In absence of a direct measurement of $Z$, the bias between $X$ and $Z$ (i.e., $\alpha_{X|Z}\neq0$) is degenerate with the estimated overall normalization of the scaling between $Y$ and $Z$. The regression  can only constrain the relative bias between $X$ and $Y$ \citepalias{se+et15_comalit_I}. 

By default, \textsc{LIRA} uses a single Gaussian distribution to model the intrinsic distribution of the independent variable ($n_\text{mix}=1$). For mixtures, \textsc{LIRA} adopts a Dirichlet distribution for the probability coefficients\footnote{In \textsc{LIRA}, the shortcut for this prior is \texttt{ddirch}.} \citep{kel07}
\beq
\label{eq_pri_5}
\pi_1, ..., \pi_{n_\text{mix}} \sim \text{Dirichlet}(1, ..., 1),
\eeq
which is equivalent to a uniform prior under the constraint  $\sum_{k=1}^{n_\text{mix}}\pi_k =1$. The number of mixture components $n_\text{mix}$ has to be fixed. Alternative approaches can determine the optimal number of Gaussian components modeling the intrinsic distribution through a Dirichlet process \citep{man15}.

By default, the regression adopts linear models with no breaks. There is no knee and the slope $\beta_{Y|Z,\text{knee}}$ and tilt $\delta_{Y|Z,\text{knee}}$ tally with $\beta_{Y|Z}$ and $\delta_{Y|Z}$, respectively. Otherwise, a flat prior can be adopted for $Z_\text{knee}$ and a Student's-t prior for $\beta_{Y|Z,\text{knee}}$ and $\delta_{Y|Z,\text{knee}}$, when applicable. The transition length is set by default to $l_\text{knee}=10^{-4}$, which makes the transition abrupt.

The above considerations drove the choice of the default priors listed in Table~\ref{tab_par}. For current data-sets, the $\gamma_{D}$ parameters can be set safely to zero in any regression, see App.~\ref{app_reds}. The only exception is $\gamma_{\mu_Z,D}$, which is crucial to model the time-evolution of the intrinsic function of flux-selected sample \citepalias{se+et15_comalit_IV}.

\section{Simulations}
\label{sec_simu}

I investigated how the approach detailed in Sec.~\ref{sec_regr} can recover scaling relations in presence of noise, scatter, and selection biases. The approach was tested with toy models and simulated data sample. I set up a basic scheme, whose essential features were as follows.

The independent variables $Z$ were drawn from a normal distribution with mean $\mu_{Z,0}=0$, and standard deviation $\sigma_{Z,0}=0.3$. The values of $Y$ were simulated assuming $\alpha_{Y|Z}=0$, $\beta_{Y|Z}=1$ and $\sigma_{Y|Z,0}=0.1$. $X$ tallies with $Z$ ($\alpha_{X|Z}=0$, $\beta_{X|Z}=1$ and $\sigma_{X|Z,0}=0$). All other parameters were set to zero.

The measurement errors were different for each data point. The variances in the measurement errors, $\delta x^2$ and $\delta y^2$  were drawn from a scaled inverse $\chi^2$-distribution with 5 degrees of freedom \citep{kel07}. The scale parameters, which dictate the typical size of the measurement errors were set to $0.1^2$. I simulated a varying degree of correlations among the measurement errors. The correlations were drawn from a uniform distribution ranging from 0.0 to 0.4.

In case of a sample covering a redshift range, the above parameters values were intended as the normalizations at the reference redshift, $z_\text{ref}=0.01$. The default time evolution was set to  $\gamma_{Y|Z}=1$ and the redshifts were drawn from a lognormal distribution. I considered $F_z=E_z$ as time factor and I computed cosmological distances as angular diameter distances.

For each case study I generated $10^3$ data sets, each one with $n_\text{sample}=100$ data points, as typical of current samples \citepalias{se+et15_comalit_IV}. The scaling relations, the scatters, and the intrinsic $Z$-distributions were recovered with \textsc{LIRA}. Parameter priors were set to the default distributions listed in Table~\ref{tab_par}. Posterior probability distributions were constrained with Markov chains generated with a Gibbs sampler. The \textsc{LIRA} package relies on the \textsc{JAGS} (Just Another Gibbs sampler) library to perform the sampling.

For each data set, I computed the parameter medians from the chains and I studied the distributions of the medians of the ensemble.

The simulation scheme was modified and made more complex if needed to highlight some aspects. On occasion, I simulated a skewed and evolving intrinsic distribution of the independent variable, scattered values of $X$, and time evolving scatter or slope.

When applicable, I also considered other publicly available methods such as \textsc{BCES}\footnote{\url{http://www.astro.wisc.edu/~mab/archive/stats/stats.html}} \citep{ak+be96} and \textsc{LINMIX}\footnote{\url{http://idlastro.gsfc.nasa.gov/ftp/pro/math/linmix_err.pro}} or its generalization to multivariate regression \textsc{MLINMIX}\footnote{\url{http://idlastro.gsfc.nasa.gov/ftp/pro/math/mlinmix_err.pro}} \citep{kel07}. The underlying hypotheses of these methods are well known and I only used them when applicable. I could not consider \textsc{BCES} for time-dependent populations or \textsc{MLINMIX} for time evolving scatters, Malmquist biased samples or in case of deviation from linearity.

\subsection{Skewed distribution}
\label{sec_z_ref}

\begin{table}
\caption{Scaling parameters recovered from samples whose independent variable follows a skewed distribution. I report the bi-weight estimators of the distribution of the median values of the simulated chains.
}
\label{tab_sim_skew}
\centering
\resizebox{\hsize}{!} {
\begin{tabular}[c]{l  l  r@{$\,\pm\,$}l  r@{$\,\pm\,$}l  r@{$\,\pm\,$}l  r@{$\,\pm\,$}l }
\hline
\noalign{\smallskip} 
	parameter & input & \multicolumn{2}{c}{\textsc{LIRA}} & \multicolumn{2}{c}{\textsc{LIRA}}& \multicolumn{2}{c}{\textsc{MLINMIX}} & \multicolumn{2}{c}{\textsc{BCES}}  \\ 
	\multicolumn{2}{c}{} & \multicolumn{2}{c}{$n_\text{mix}=1$}& \multicolumn{2}{c}{$n_\text{mix}=3$}&  \multicolumn{2}{c}{$n_\text{mix}=3$} & \multicolumn{2}{c}{}  \\ 
\noalign{\smallskip} 
\hline
	 \multicolumn{10}{c}{$z=z_\text{ref}$}  \\
	 $\alpha_{Y|Z}$ &  $[0]$& 0.00&0.03&  0.00&0.03& 0.00&0.03 &0.00&0.04	\\
	 $\beta_{Y|Z}$ &  $[1]$& 1.02&0.12&  1.01&0.11& 1.00&0.11 &1.02&0.15	\\
	  $\sigma_{Y|Z,0}$ &  $[0.1]$& 0.09&0.03&  0.09&0.03& 0.10&0.02 &0.13&0.01	\\
\noalign{\smallskip} 
	    \multicolumn{10}{c}{redshift evolution}  \\
	  \multicolumn{2}{c}{} & \multicolumn{2}{c}{}& \multicolumn{2}{c}{}&  \multicolumn{2}{c}{$n_\text{mix}=1$} & \multicolumn{2}{c}{}  \\ 
	  $\alpha_{Y|Z}$    &   [0]    & -0.01&0.10&   \multicolumn{2}{c}{}& 0.01&0.09 & \multicolumn{2}{c}{}	\\
	 $\beta_{Y|Z}$       &   [1]    & 1.02&0.12&   \multicolumn{2}{c}{}& 0.99&0.11 & \multicolumn{2}{c}{}	\\
	 $\gamma_{Y|Z}$  &   [1]    & 0.83&0.45&   \multicolumn{2}{c}{}& 1.02&0.47 & \multicolumn{2}{c}{}	\\
	  $\sigma_{Y|Z,0}$ &   [0.1] & 0.09&0.03&   \multicolumn{2}{c}{}& 0.10&0.02 & \multicolumn{2}{c}{}	\\
	  \hline
	\end{tabular}
	}
\end{table}

The accurate modeling of the intrinsic distribution of the covariate variable is crucial to unbiased linear regression. To test its effect, I considered an asymmetric distribution. I modified the basic simulation scheme by drawing $Z$ from a skew-normal distribution with shape parameter $\alpha_{Z,0,\text{skew}}=3.0$. The location and the scale parameter were made to coincide with the mean and the standard deviation of the basic normal distribution. Results are reported in Table~\ref{tab_sim_skew}.

\subsubsection{No time evolution}
\label{sec_sim_skew}

\begin{figure}
\begin{tabular}{c}
\resizebox{\hsize}{!}{\includegraphics{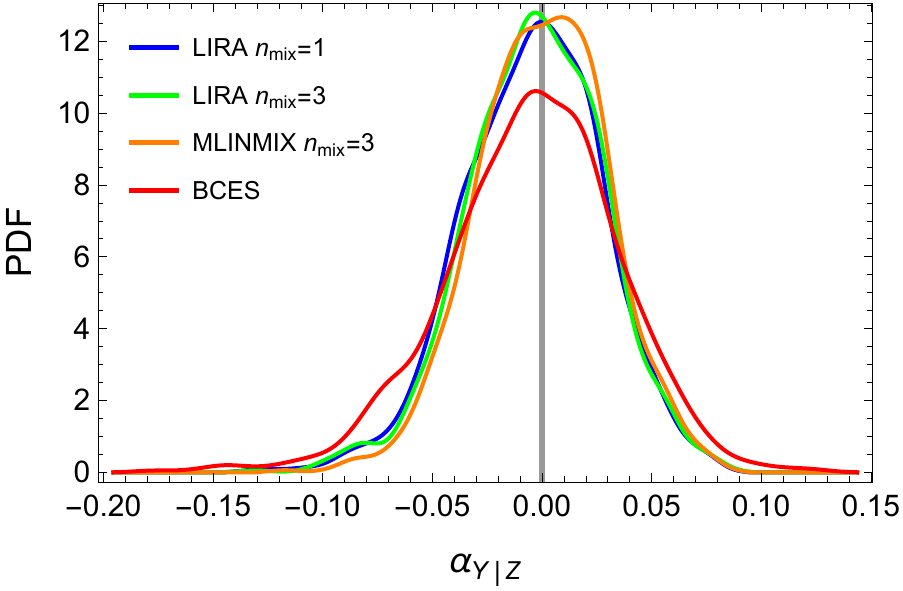}} \\
\resizebox{\hsize}{!}{\includegraphics{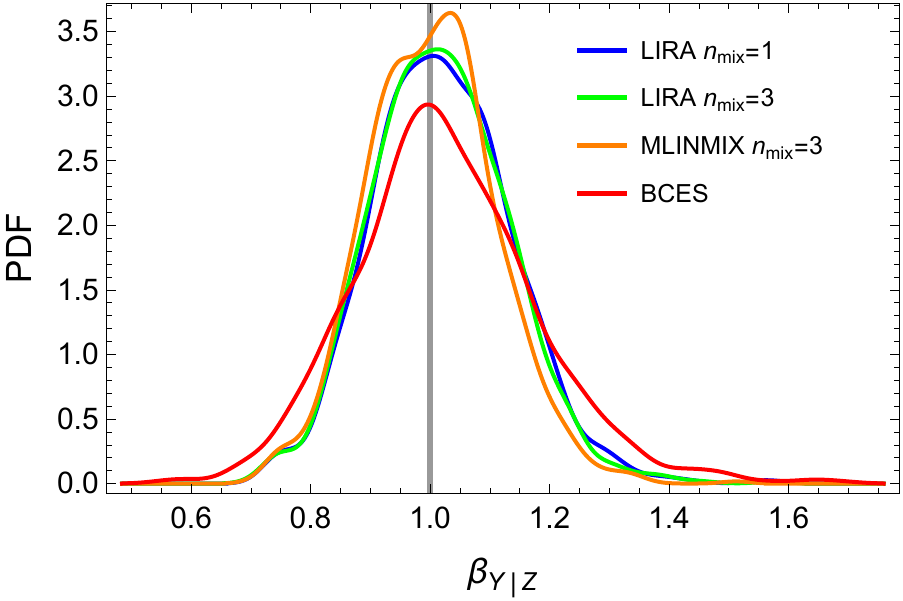}} \\ 
\resizebox{\hsize}{!}{\includegraphics{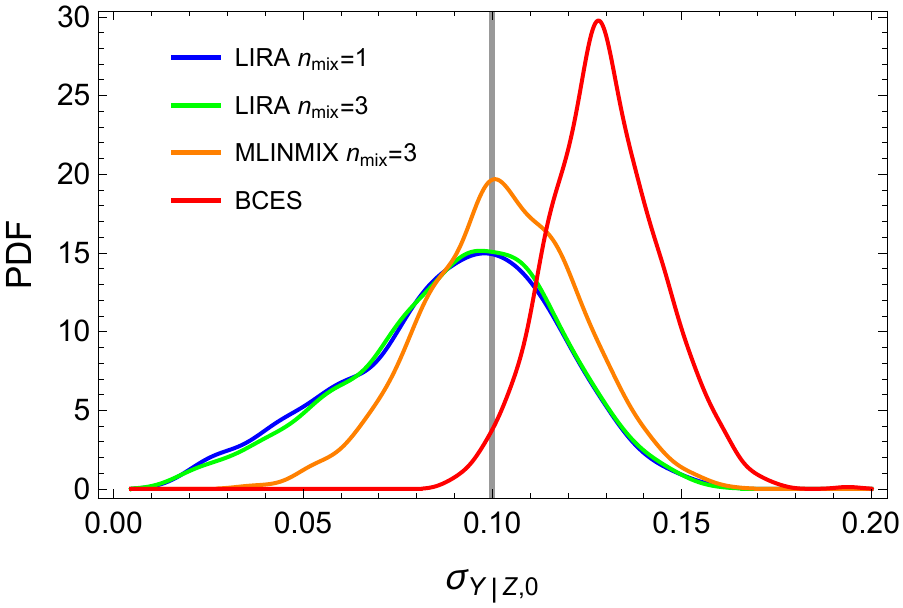}} \\
\end{tabular}
\caption{Distributions of the median parameters of the scaling relation of samples drawn from a skewed intrinsic distribution $p(Z)$. The blue, green, orange and red lines are the smoothed histograms of the distributions obtained with \textsc{LIRA} by modeling $p(Z)$ with a single Gaussian function, with \textsc{LIRA} by adopting a mixture of 3 Gaussian functions, with \textsc{LINMIX} by adopting a mixture of 3 Gaussian functions, and with \textsc{BCES}, respectively. The vertical gray lines are set at the input parameters. From the top to the bottom panel: the intercept, the slope, and the intrinsic scatter.}
\label{fig_pdf_skewed_zRef}
\end{figure}

I first considered samples drawn at the same reference redshift. Results are summarized in Table~\ref{tab_sim_skew} and Fig.~\ref{fig_pdf_skewed_zRef}. To recover the parameters, I considered either a simple \textsc{LIRA} model with just one normal distribution to shape $p(Z)$ or a mixture of three components. For comparison, I also computed parameter chains with \textsc{LINMIX} adopting a mixture of three Gaussian distributions and the \textsc{BCES}(Y|X) estimator. The original work introducing \textsc{BCES} did not advocate any method to compute the intrinsic scatter, which I computed following \citet{pra+al09}. 

Input parameters are well reproduced by all methods. In this setting, the agreement between \textsc{LIRA} and \textsc{LINMIX} is excellent. This is expected since the main assumptions of the two methods are equivalent. Minor differences come from the different choice of the priors, which are of lesser importance when the data analysis is dominated by the likelihood and by the data. The parameter distributions agree very well, even though the distribution of the intrinsic scatter $\sigma_{Y|Z,0}$ from \textsc{LIRA} has a more pronounced tail at small values. This tail is not present in richer data-sets, see Sec.~\ref{sec_sim_scat}.

\textsc{BCES} recovers well the central values of the slope and of the intercept but statistical uncertainties are larger. The intrinsic scatter estimate is biased high.

Even though the intrinsic simulated distribution of $Z$ is skewed, there is no real improvement by augmenting the number of mixture components, see Table~\ref{tab_sim_skew}. As far as the intrinsic distribution is unimodal and the sample is not too rich, one Gaussian component is enough to recover the regression parameters (\citealt{kel07}, \citetalias{ser+al15_comalit_II}).

\subsubsection{Evolution with redshift}
\label{sec_sim_skew_z}

\begin{figure}
\begin{tabular}{c}
\resizebox{0.45\textwidth}{!}{\includegraphics{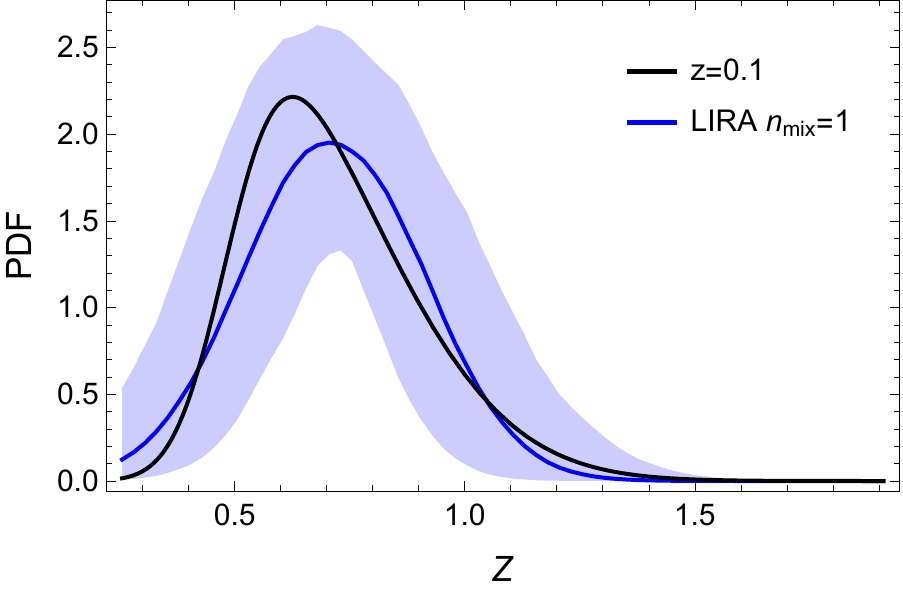}} \\
\resizebox{0.45\textwidth}{!}{\includegraphics{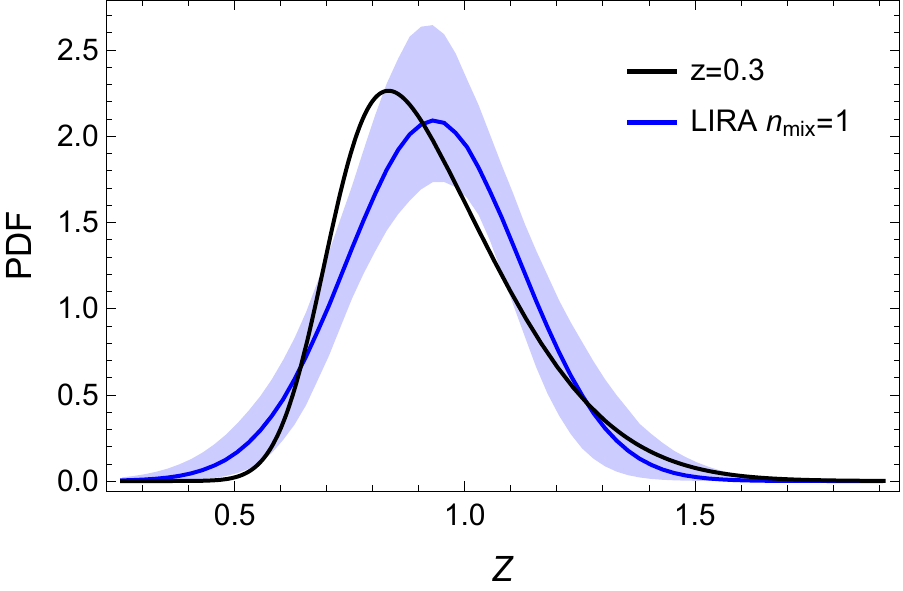}} \\ 
\resizebox{0.45\textwidth}{!}{\includegraphics{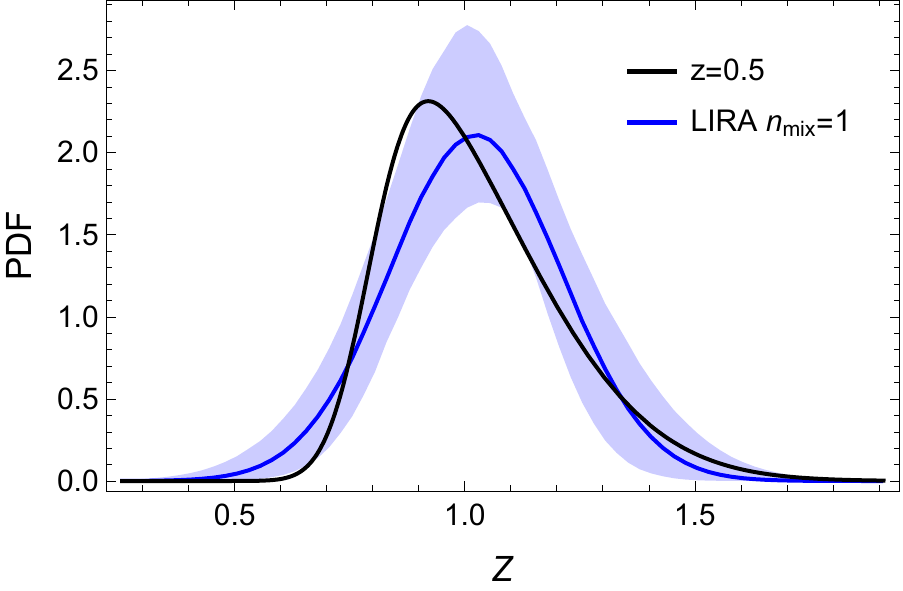}} \\
\resizebox{0.45\textwidth}{!}{\includegraphics{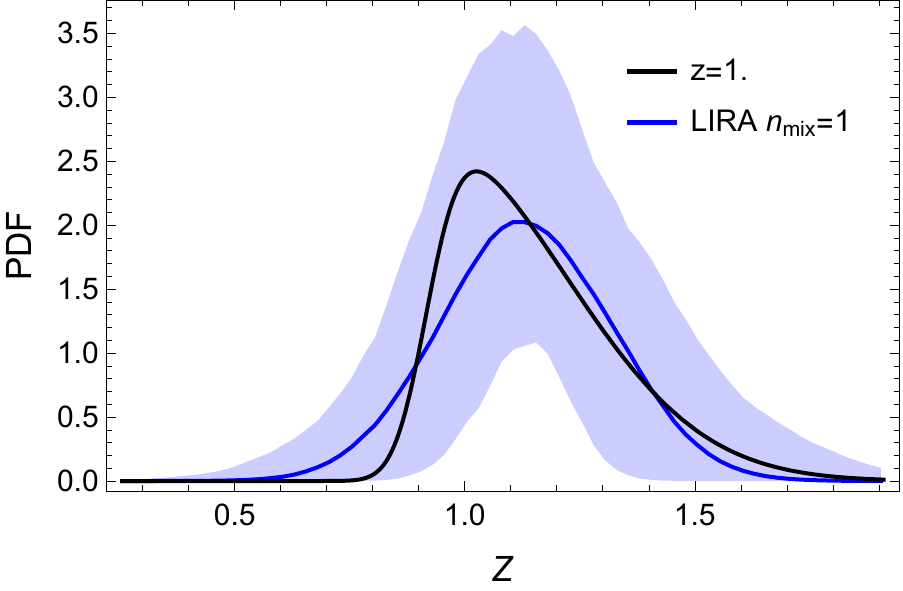}} \\
\end{tabular}
\caption{The reconstructed intrinsic distribution of the independent variable $Z$ at different redshifts. From top to the bottom, $z=0.1$, 0.3, 0.5, 1.0. The black line is the input distribution, the blue line is the median reconstructed relation, the shadowed blue region encloses the 1-$\sigma$ confidence region at a given $Z$. For a total of $n_\text{sample}=100$ data, we expect $\sim 21$, 51, 20 and 1 sources in the redshift range $0.0\le z\le 0.2$, $0.2\le z\le 0.4$, $0.4\le z\le 0.6$ and $0.9\le z\le 1.1$, respectively.}
\label{fig_pZ_skewed_z}
\end{figure}

\begin{figure}
\begin{tabular}{c}
\resizebox{0.44\textwidth}{!}{\includegraphics{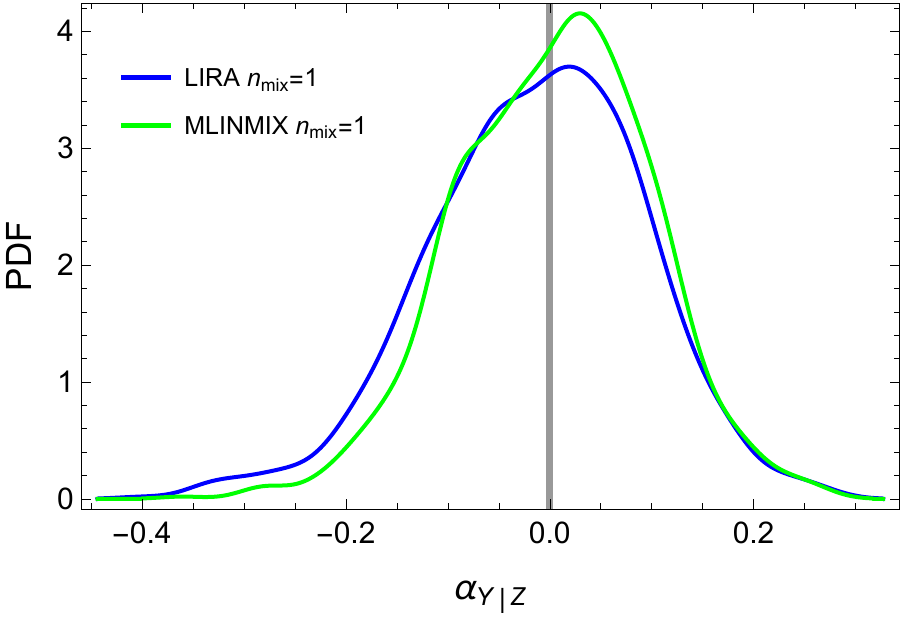}} \\
\resizebox{0.44\textwidth}{!}{\includegraphics{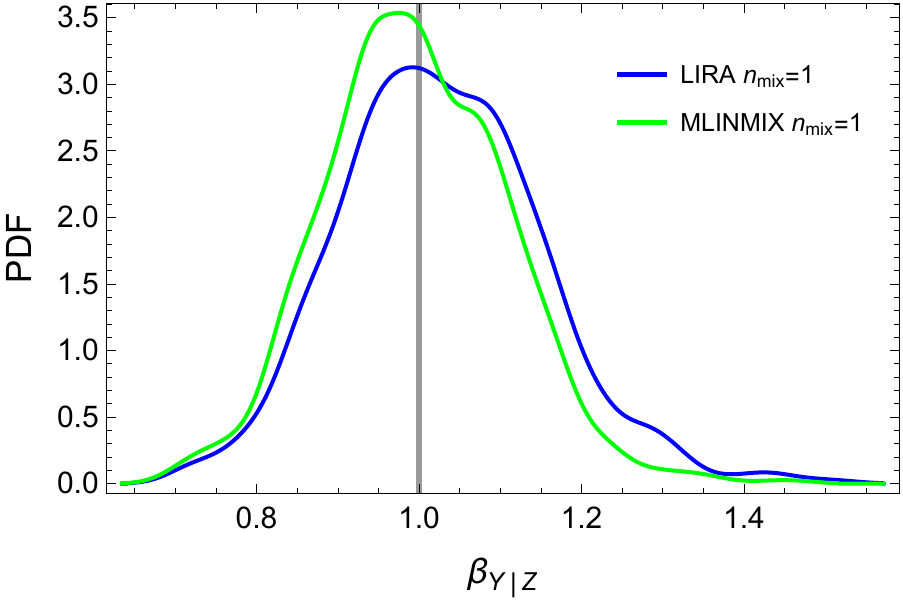}} \\ 
\resizebox{0.44\textwidth}{!}{\includegraphics{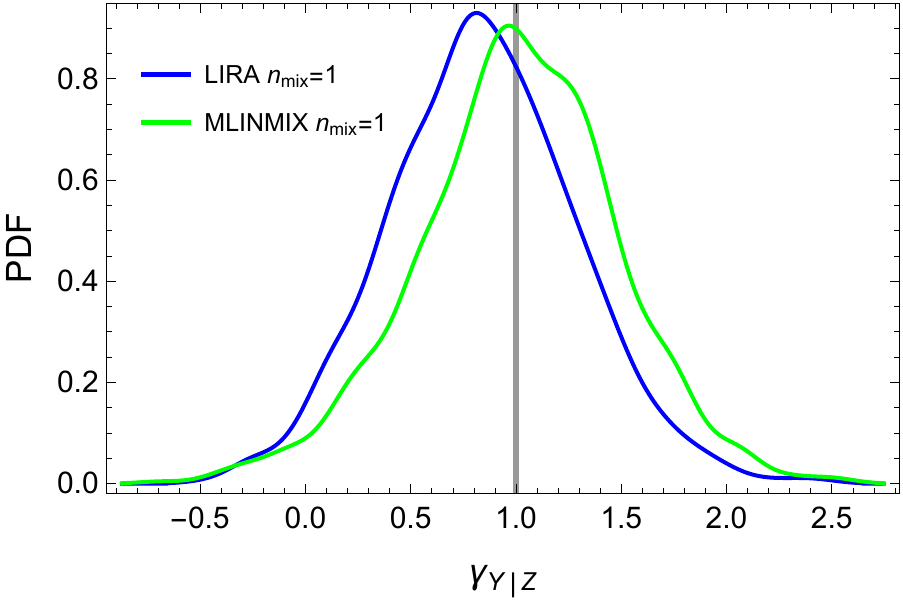}} \\ 
\resizebox{0.44\textwidth}{!}{\includegraphics{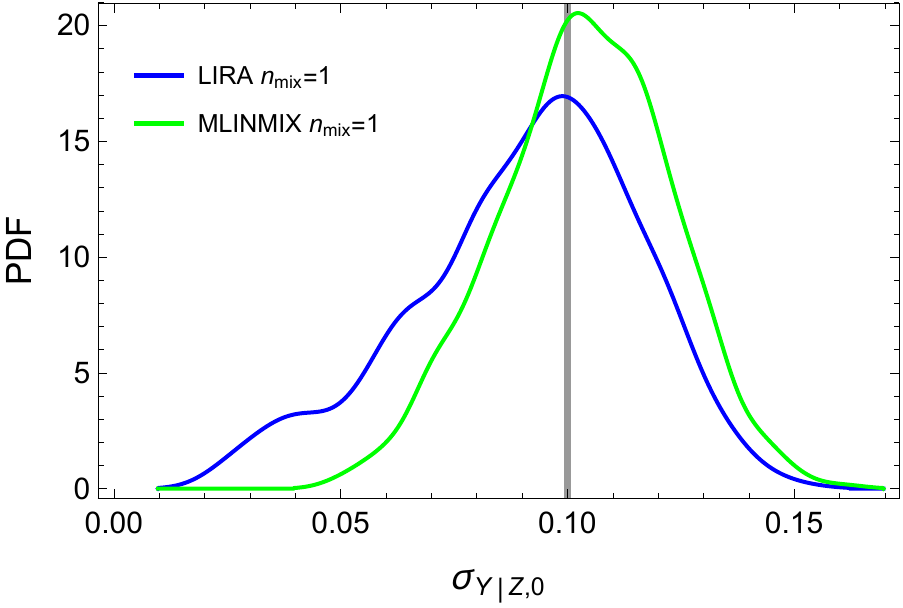}} \\
\end{tabular}
\caption{Distributions of the median parameters obtained from samples with a skewed and time-evolving intrinsic distribution $p(Z,z)$. The blue (green) lines are the smoothed histograms of the distributions obtained with \textsc{LIRA} (\textsc{M\textsc{LINMIX}}). $p(Z,z)$ was modeled with a single Gaussian function. The vertical gray lines are set at the input parameters. From the top to the bottom panel: the intercept, the slope, the time evolution, and the intrinsic scatter.}
\label{fig_pdf_skewed_z}
\end{figure}

I then considered samples covering an extended redshift range. Redshifts were drawn from a lognormal distribution such that $\ln z$ has mean $\ln (0.3)$ and standard deviation 0.5. In these simulations, the skewed distribution of the independent variable is time evolving. The location parameter of the input distribution evolves with the redshift as in Eq.~(\ref{eq_mix_2}) with $\gamma_{\mu_Z,F_z}=0.5$  and $\gamma_{\mu_Z,D}=0.5$; the scale parameter is fixed, whereas the shape parameter evolves with redshift proportionally to $E_z$. The input intrinsic scatter $\sigma_{Y|Z}$ is redshift independent.

I recovered the input parameters by modeling $p(Z)$ with a single normal distribution whose mean and standard deviation evolve with time. The prior on $\gamma_{\sigma_Z,F_z}$ was set such that the inverse variance follows a Gamma distribution, see Eq.~(\ref{eq_pri_3}), whereas $\gamma_{\sigma_Z,D}$ was set to zero. Even if the input $p(Z)$ distribution is asymmetric, this modeling is enough to recover the time evolution of $p(Z)$, see Figure~\ref{fig_pZ_skewed_z}, and to get unbiased values of the scaling parameters, see Table~\ref{tab_sim_skew} and Figure~\ref{fig_pdf_skewed_z}.

For comparison I performed the regression with \textsc{MLINMIX} too. The \textsc{LIRA} scheme differs from the multivariate analysis detailed in \citet{kel07} in one major feature. \textsc{LIRA} models the intrinsic distribution with a mixture of one-dimensional Gaussian components whose means and standard deviations are time-dependent. On the other hand, \textsc{MLINMIX} models the bi-dimensional distribution of $Z$ and $T$ with a mixture of bi-dimensional Gaussian components whose means and variances are not time-evolving. Notwithstanding this important difference and some minor differences due to the prior choice, both approaches can recover with good accuracy the scaling parameters, see Figure~\ref{fig_pdf_skewed_z}.

As far as the scaling parameters and the scatter is concerned, it is important to model the non-uniformity of the distribution of the intrinsic distribution. Details on the exact form of the distribution are of second order. The safer approach to model noisy ad sparse samples is to use the simplest model, e.g., a single normal distribution for $p(Z)$. In samples of order of one hundred of objects, there are just a few items at high redshift. Enforcing a more complex distribution, such as a skewed Gaussian, to model sparse data can bias the results towards a few outliers due to overfitting. Complex distributions are recommended only for very rich samples.

\subsection{Eddington bias}
\label{sec_sim_eddi}

\begin{table}
\caption{Scaling parameters recovered from biased samples. For each case (listed in Col.~1), I reported on consecutive rows the values of the parameters obtained with regressions which do either correct or not correct for the bias. Reported values are the bi-weight estimators of the distribution of the median values of the simulated chains.
}
\label{tab_sim_bias}
\centering
\resizebox{\hsize}{!} {
\begin{tabular}[c]{l  r@{$\,\pm\,$}l  r@{$\,\pm\,$}l  r@{$\,\pm\,$}l }
\hline
\noalign{\smallskip} 
	case  & \multicolumn{2}{c}{ $\alpha_{Y|Z}$} & \multicolumn{2}{c}{$\beta_{Y|Z}$}& \multicolumn{2}{c}{$\sigma_{Y|Z,0}$}  \\
\hline
	input &  \multicolumn{2}{c}{[0]} &  \multicolumn{2}{c}{[1]} &  \multicolumn{2}{c}{[0.1]} \\
	\noalign{\smallskip} 
	\multicolumn{7}{l}{Eddington bias ($\sigma_{X|Z} \neq 0$)}  \\
	corrected  &  0.00&0.02&  1.01&0.09  & 0.09&0.03	\\
	biased      &  0.00&0.02&  0.90&0.07  & 0.14&0.02	\\
\noalign{\smallskip} 
	\multicolumn{7}{l}{Malmquist bias}  \\
	corrected  &  0.01&0.03&  1.06&0.11  & 0.09&0.03	\\
	biased      &  0.03&0.02&  0.90&0.09  & 0.08&0.03	\\ 
\noalign{\smallskip} 
	\multicolumn{7}{l}{Linearity break (knee)}  \\
	corrected  &  0.00&0.03&  0.98&0.18  & 0.10&0.04	\\
	biased      &  -0.05&0.02&  1.31&0.12  & 0.15&0.03	\\ 
	  \hline
	\end{tabular}
	}
\end{table}

\begin{figure}
\begin{tabular}{c}
\resizebox{\hsize}{!}{\includegraphics{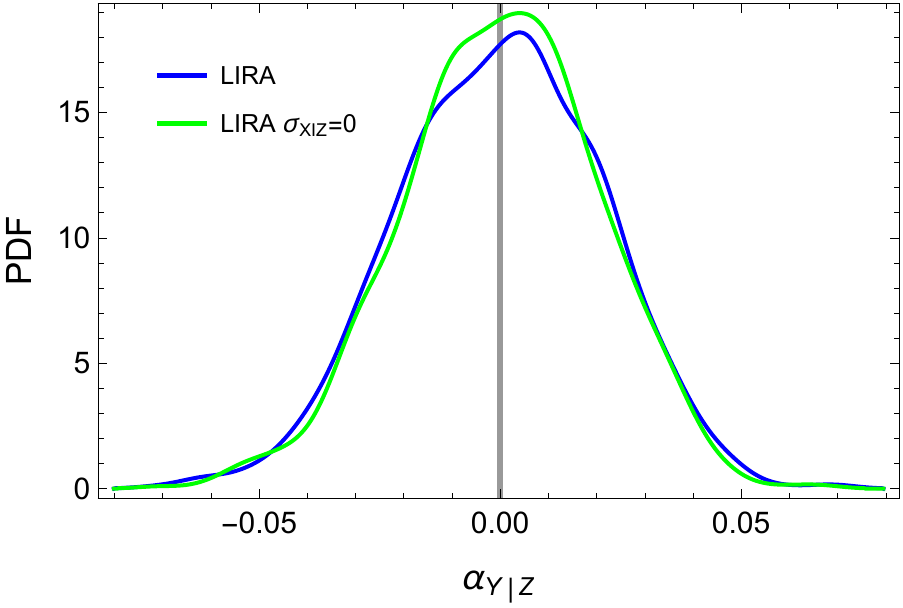}} \\
\resizebox{\hsize}{!}{\includegraphics{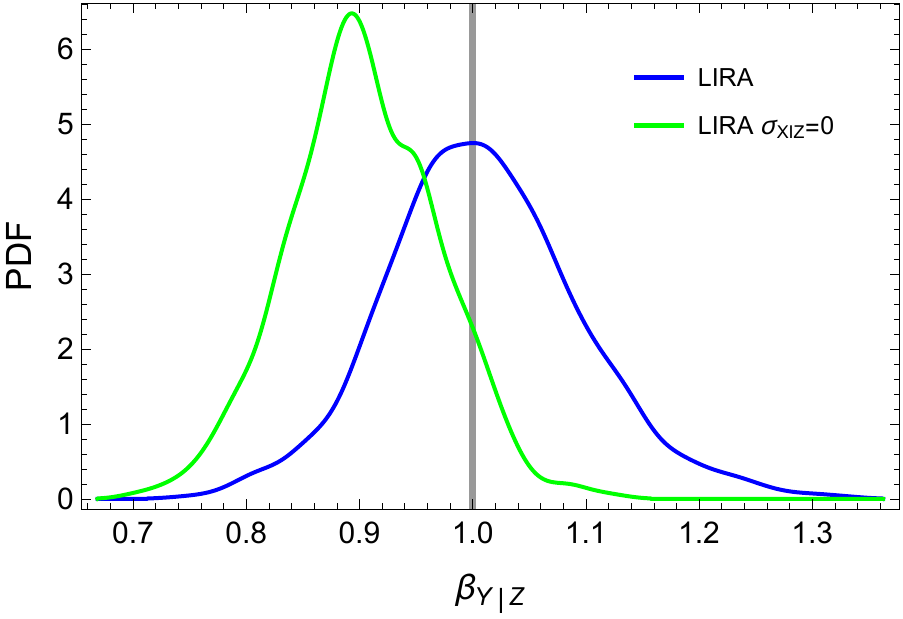}} \\ 
\resizebox{\hsize}{!}{\includegraphics{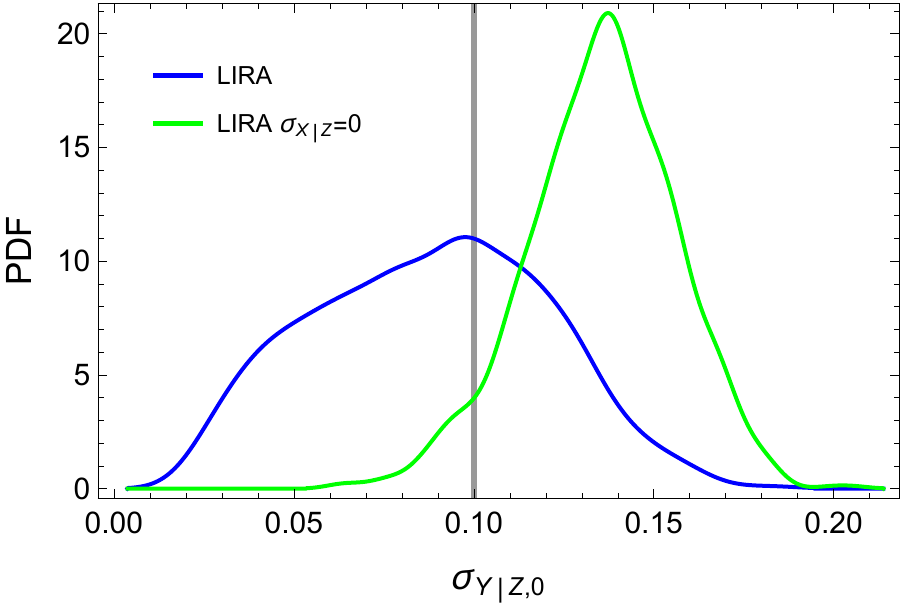}} \\
\end{tabular}
\caption{Distributions of the median parameters obtained from samples affected by Eddington bias. The blue lines are the smoothed histograms of the distributions obtained by considering the intrinsic scatter in the covariate variable. The green lines plot the results from a biased fit. The vertical gray lines are set at the input parameters. From the top to the bottom panel: the intercept, the slope, and the intrinsic scatter.}
\label{fig_pdf_sigmaXIZ}
\end{figure}

The Eddington bias can affect the estimate of the scaling parameters if the measurement errors on the covariate variable are not accounted for \citep{edd40} or the $X$ variable is a scattered proxy of the latent covariate $Z$ \citep{se+et15_comalit_I}. Here, we are mostly interested in the second case, which is often overlooked.

I simulated the samples by assuming that the measurable $X$ is an unbiased ($\alpha_{X|Z}=0$ and $\beta_{X|Z}=1$) but scattered ($\sigma_{X|Z,0}=0.1$) proxy of $Z$. The data were fitted with either a corrected procedure, where the intrinsic scatter $\sigma_{X|Z,0}$ is a model parameter, or a biased procedure with  $\sigma_{X|Z,0}=0$. Results are summarized in Table~\ref{tab_sim_bias} and Fig.~\ref{fig_pdf_sigmaXIZ}. 

The corrected procedure recovers the input parameters very well whereas systematic errors are significant if we do nor correct for the Eddington bias. The Eddington bias makes the observed relation flatter and inflates the intrinsic scatter. Since I considered a scatter $\sigma_{X|Z}$ independent of $Z$, the bias has a symmetric action and  the pivot point of the relation does not change. The normalization is not affected. Statistical uncertainties on the regression parameters are underestimated, as usual in biased measurements.

\subsection{Malmquist bias}
\label{sec_sim_malm}

\begin{figure}
\begin{tabular}{c}
\resizebox{\hsize}{!}{\includegraphics{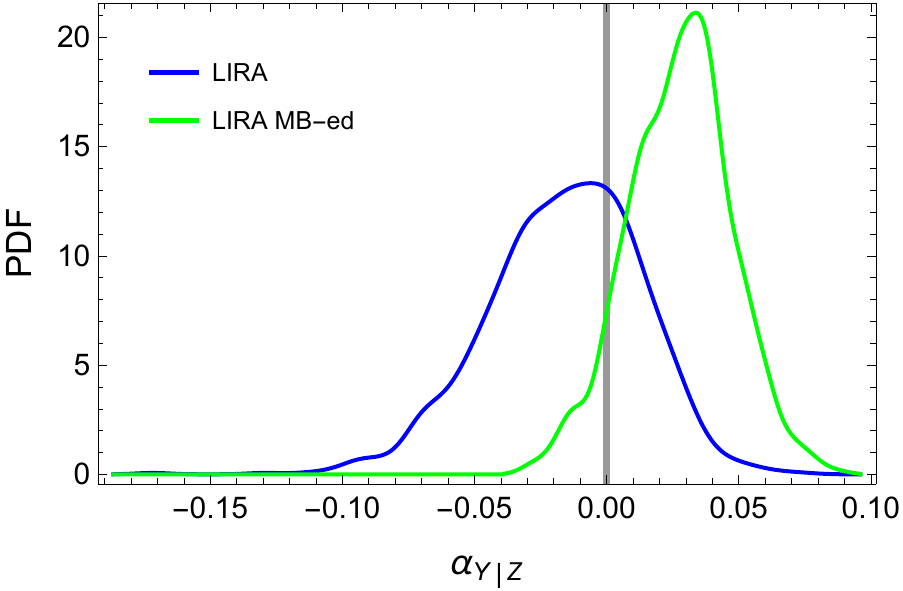}} \\
\resizebox{\hsize}{!}{\includegraphics{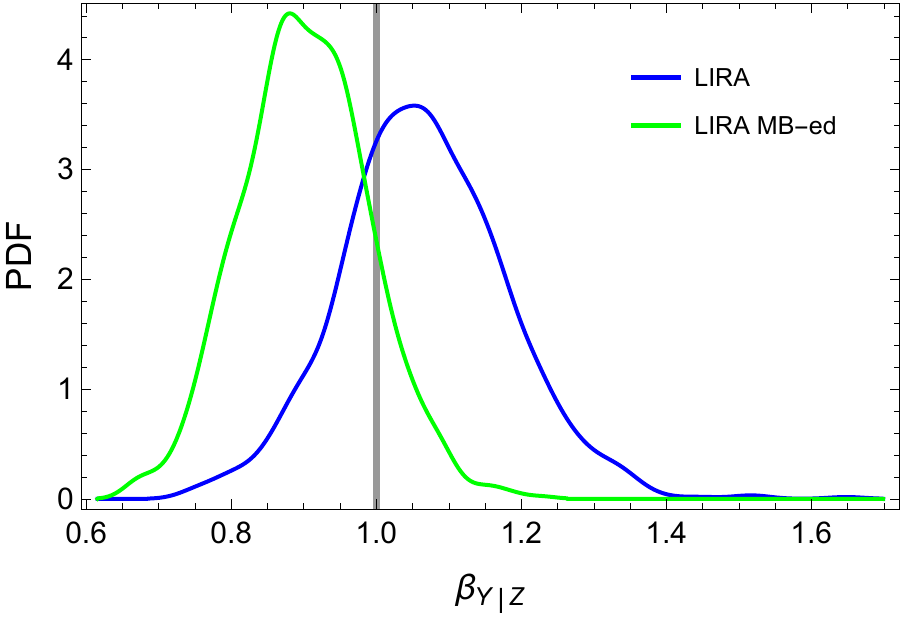}} \\ 
\resizebox{\hsize}{!}{\includegraphics{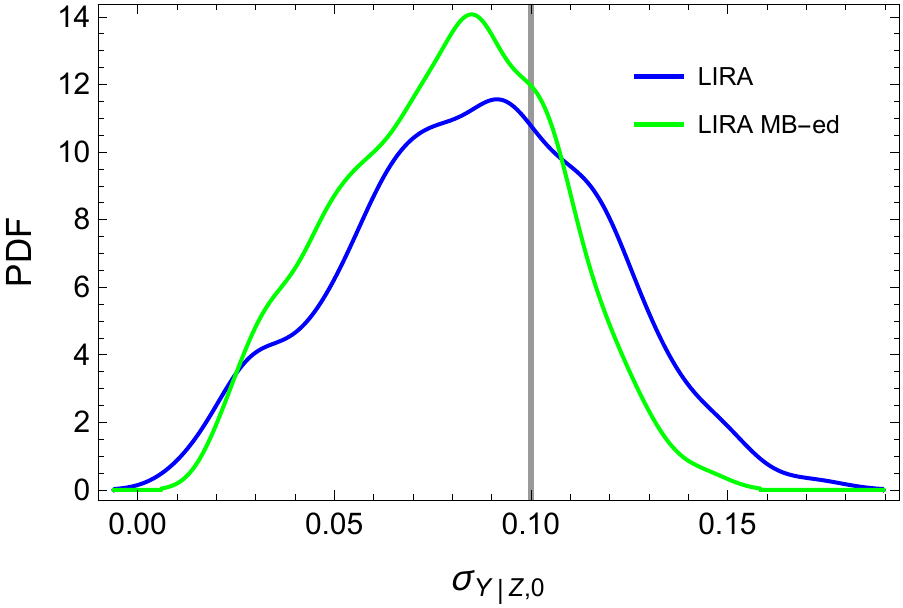}} \\
\end{tabular}
\caption{Distributions of the median parameters obtained from samples selected in the response variable. The blue lines are the smoothed histograms of the distributions obtained by correcting for the Malmquist bias. The green lines plot the results from a biased fit. The vertical gray lines are set at the input parameters. From the top to the bottom panel: the intercept, the slope, and the intrinsic scatter.}
\label{fig_pdf_MB}
\end{figure}

The Malmquist bias has long been known \citep{mal20}. Still, it can be difficult to tackle. Proposed recipes consider the correction of the measured values of individual objects, which needs a guess on the intrinsic scatter, or the modeling through a proper definition of the selection efficiency in the likelihood function \citep{vik+al09}.

To test the effect of the bias, I simulated the samples as in the standard case but I only kept objects whose measured response exceeded a threshold value ($y>y_\text{th}=-0.3$). Nearly 80 per cent of the items makes the cut. Results are summarized in Table~\ref{tab_sim_bias} and Fig.~\ref{fig_pdf_MB}. 

The Malmquist bias makes the observed relation flatter. If the bias is not corrected for, the measured slope is biased toward zero whereas the measured intercept is biased high. The scatter is affected too. It can be underestimated.

\subsection{Linearity break}
\label{sec_sim_brok}

\begin{figure}
\begin{tabular}{c}
\resizebox{\hsize}{!}{\includegraphics{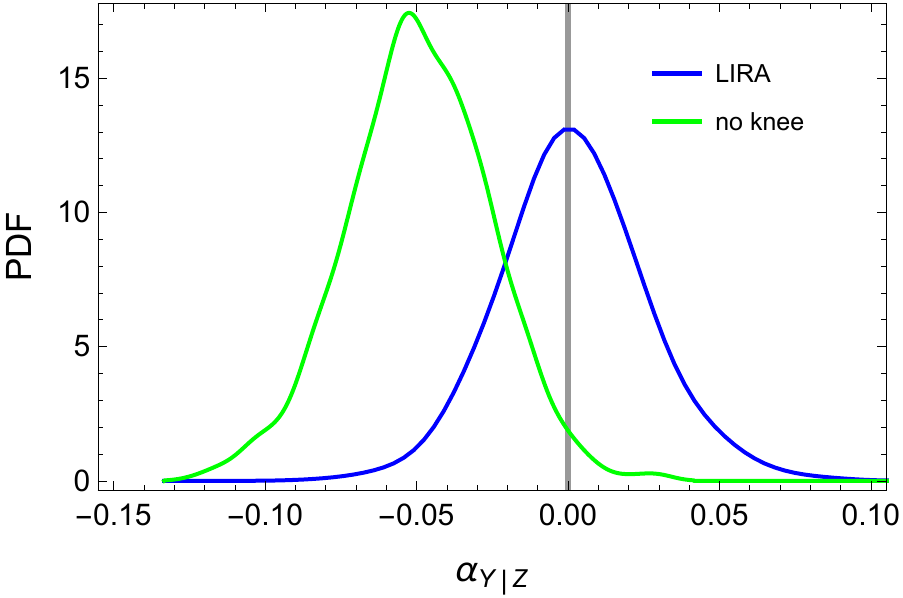}} \\
\resizebox{\hsize}{!}{\includegraphics{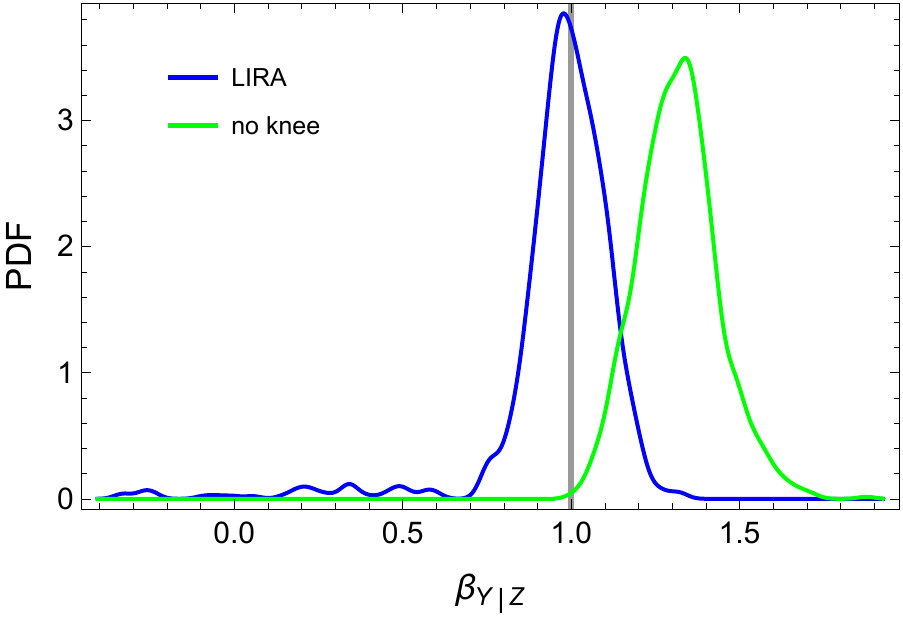}} \\ 
\resizebox{\hsize}{!}{\includegraphics{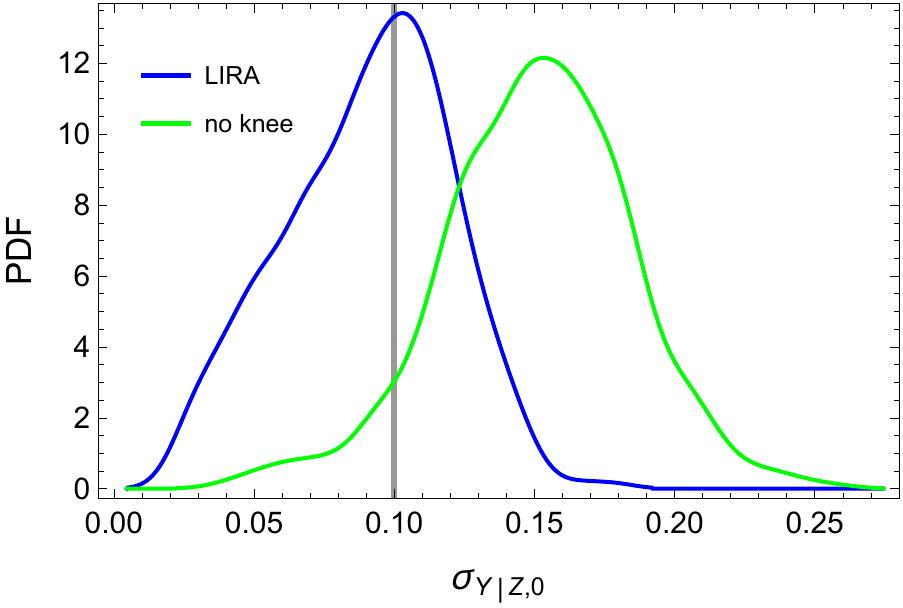}} \\
\end{tabular}
\caption{Distributions of the median parameters of a broken power-law. The blue lines are the smoothed histograms of the distributions obtained by fitting the simulated data with a scattered broken power law. The green lines plot the results from a biased linear fit. The vertical gray lines are set at the input parameters. From the top to the bottom panel: the intercept, the slope, and the intrinsic scatter.}
\label{fig_pdf_knee}
\end{figure}

Different physical processes can break the linearity of a scaling relation. In the formation and evolution of galaxy clusters, baryonic and energetic effects are relevant in small objects and can challenge the dominance of the gravitational force. A bent scaling relation can be more apt to model the process.

I simulated a broken power law relation. I set the knee at $Z_\text{knee}=\mu_{Z,0}-\sigma_{Z,0}$, i.e., $\sim16$ per cent of the sources at $Z<Z_\text{knee}$ follow a different scaling. The slope before the break was set at $\beta_{Y|Z,\text{knee}}=3.0$. Results are summarized in Table~\ref{tab_sim_bias} and Fig.~\ref{fig_pdf_knee}.

Parameter estimates obtained with a simple linear model are severely biased. The model without knee cannot distinguish the two regimes and the measured slope is a weighted average of the two real slopes. Being the slope before the knee steeper in the simulation, the intercept estimated by the linear model is biased low. If not modeled, the knee strongly affects the estimated scatter. To mimic the break and the steeper slope, the estimated scatter is severely overestimated.

\subsection{Time dependent intrinsic scatter}
\label{sec_sim_scat}

\begin{table}
\caption{Scaling parameters recovered from time evolving samples. I report the bi-weight estimators of the distribution of the median values of the simulated chains.
}
\label{tab_sim_time}
\centering
%\resizebox{\hsize}{!} {
\begin{tabular}[c]{l  l  r@{$\,\pm\,$}l  r@{$\,\pm\,$}l}
\hline
\noalign{\smallskip} 
	parameter  &input&  \multicolumn{2}{c}{unbiased} & \multicolumn{2}{c}{biased}  \\ 
\noalign{\smallskip} 
\hline
	 \multicolumn{6}{c}{Time evolving scatter}  \\
	 $\alpha_{Y|Z}$    &  [0]& 0.00&0.02&  0.00&0.02 	\\
	 $\beta_{Y|Z}$      &  [1]& 1.00&0.03&  1.00&0.03	\\
	 $\gamma_{Y|Z}$ &  [1]& 1.00&0.07&  1.00&0.08	\\
	 $\sigma_{Y|Z,0}$&  [0.1]& 0.100&0.008&  0.123&0.007	\\
	 $\gamma_{\sigma_{Y|Z,F_z}}$&  [0.5]& 0.048&0.15&   \multicolumn{2}{c}{[0]}	\\
	  \multicolumn{6}{c}{Time evolving slope}  \\
	 $\alpha_{Y|Z}$    &  [0]& 0.00&0.05&  -0.05&0.04 	\\
	 $\beta_{Y|Z}$      &  [1]& 1.00&0.07&  1.07&0.05	\\
	 $\gamma_{Y|Z}$ &  [1]& 0.93&0.59&  1.72&0.34	\\
	 $\delta_{Y|Z}$     &  [1]&  0.93&0.67 & \multicolumn{2}{c}{[0]}	\\
\noalign{\smallskip} 
	  \hline
	\end{tabular}
%	}
\end{table}

\begin{figure}
\centering
\begin{tabular}{c}
\resizebox{0.35\textwidth}{!}{\includegraphics{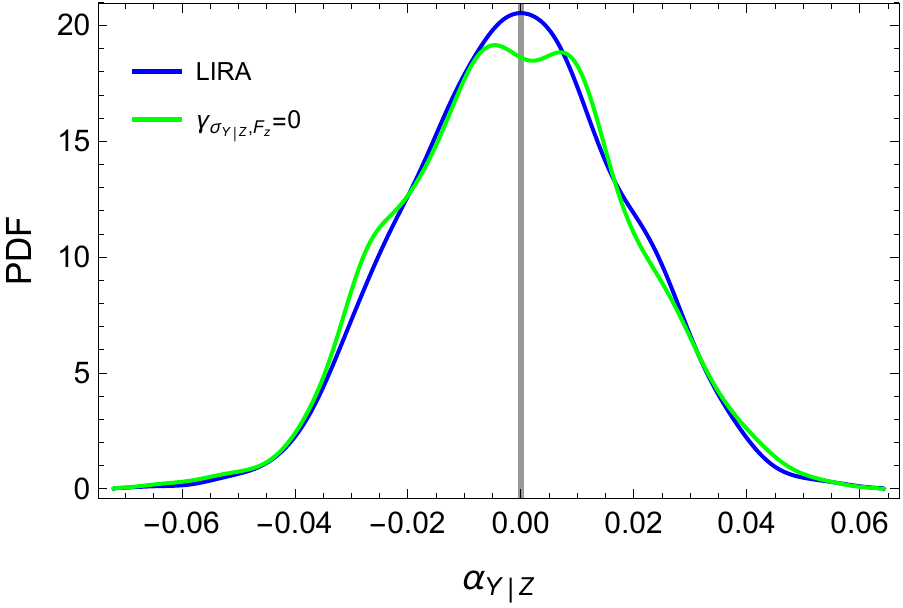}} \\
\resizebox{0.35\textwidth}{!}{\includegraphics{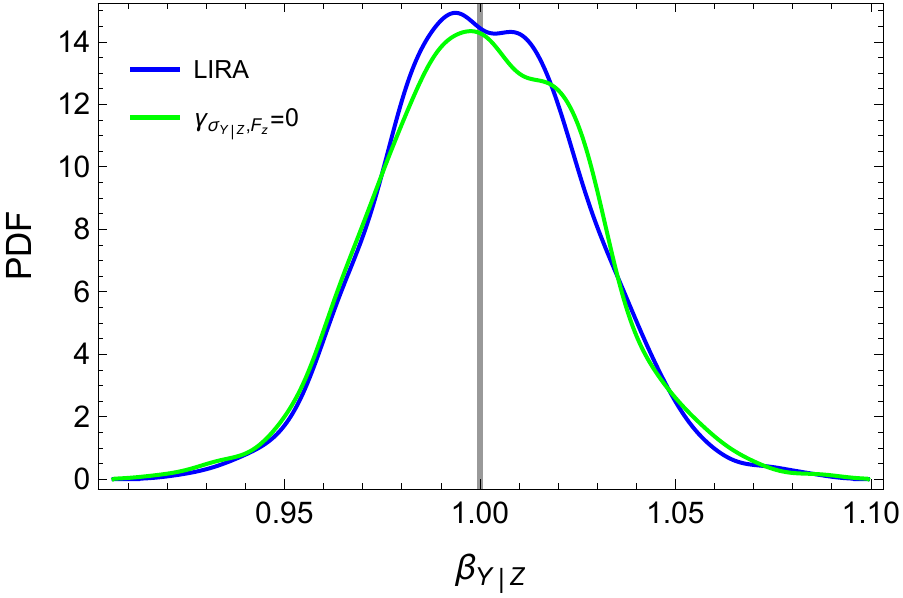}} \\ 
\resizebox{0.35\textwidth}{!}{\includegraphics{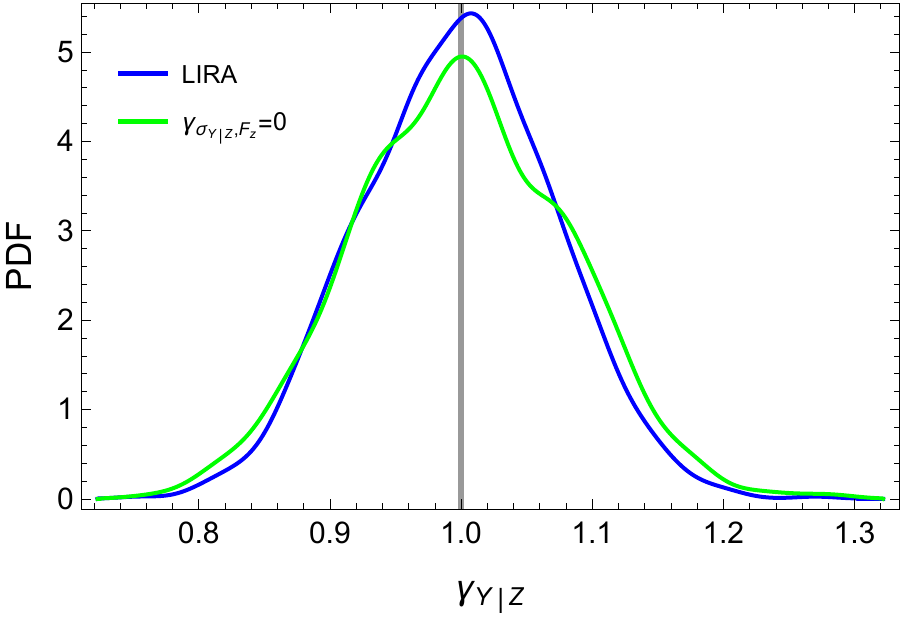}} \\ 
\resizebox{0.35\textwidth}{!}{\includegraphics{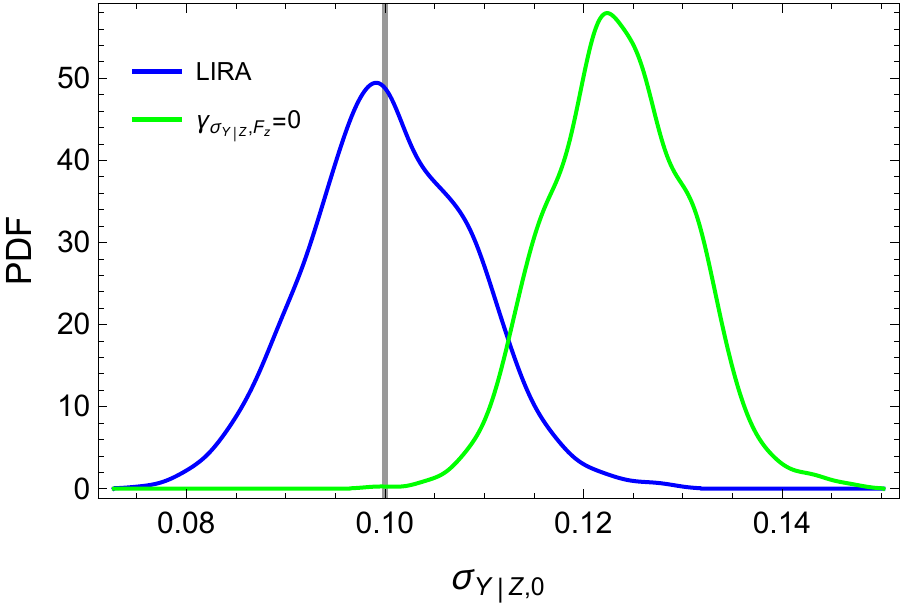}} \\
\resizebox{0.35\textwidth}{!}{\includegraphics{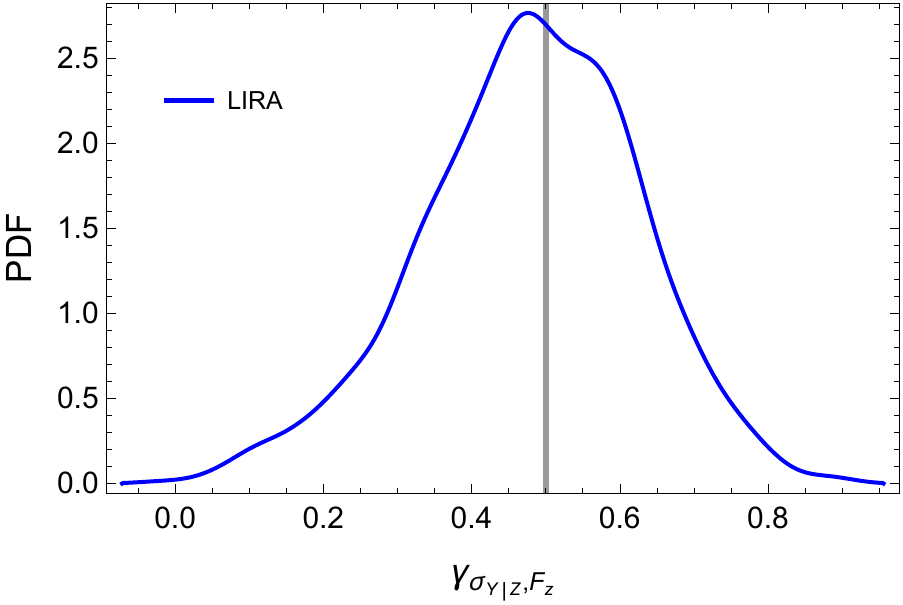}}
\end{tabular}
\caption{Distributions of the median parameters obtained from samples with time-evolving intrinsic scatter. The blue line are the smoothed histograms of the distributions obtained by fitting the simulated data with a time-dependent scatter. The green lines plot the results from a biased linear fit with $\gamma_{\sigma_{Y|Z},F_z}=0$. The vertical gray lines are set at the input parameters. From the top to the bottom panel: the intercept, the slope, the time evolution, the intrinsic scatter, and the scatter evolution.}
\label{fig_pdf_sigmaYIZz}
\end{figure}

Usual data sets are not rich enough to measure the time evolution of the intrinsic scatter \citepalias{se+et15_comalit_IV}. The $\gamma$ parameters modeling the scatter redshift dependence, i.e., $\gamma_{\sigma_{Y|Z,F_z}}$ or $\gamma_{\sigma_{Y|Z,D}}$, are better seen as noise parameters to marginalize over.

The study of the time evolution of the scatter will be at reach of future surveys \citep{eucl_lau_11}. I then increased the number of simulated sources per sample and their redshift range and I considered smaller observational errors. I simulated samples with 400 items each. The scale parameters of the scaled inverse $\chi^2$-distributions modeling the uncertainty variances $\delta x^2$ and $\delta y^2$ were set to $0.05^2$ and measurement errors were assumed to be uncorrelated. 

Redshifts were drawn from a lognormal distribution such that $\ln z$ has mean $\ln (0.5)$ and standard deviation 0.8, i.e., $\sim19$ per cent of the sources are at $z>1$. The independent variables were drawn from a time evolving normal distribution. The mean evolves with redshift as in Eq.~(\ref{eq_mix_2}) with $\gamma_{\mu_Z,F_z}=0.5$  and $\gamma_{\mu_Z,D}=0.5$; the standard deviation is constant.

 The intrinsic scatter evolved with redshift as in Eq.~(\ref{eq_bug_5}), with  $\sigma_{Y|Z,0}=0.1$, $\gamma_{\sigma_{Y|Z,F_z}}=0.5$ and $\gamma_{\sigma_{Y|Z,D}}=0$. The input intrinsic scatter at $z\sim1$ is $\sim 30$ per cent larger than the local value. The remaining parameters were set as for the other simulations.
 
Results are summarized in Table~\ref{tab_sim_time} and Fig.~\ref{fig_pdf_sigmaYIZz}. Even if we neglect the scatter evolution, the estimates of the scaling parameters are unbiased whereas the estimated intrinsic scatter is weighted over the redshift range. The corrected regression can recover both the normalization and the time evolution of the scatter. Since the simulated sample is copious and observational accuracy is improved with respect to the other simulations, the posterior distribution of the intrinsic scatter is symmetric, with no prominent tail at small values.

\subsection{Redshift dependent slope}
\label{sec_sim_tilt}

\begin{figure}
\centering
\begin{tabular}{c}
\resizebox{0.35\textwidth}{!}{\includegraphics{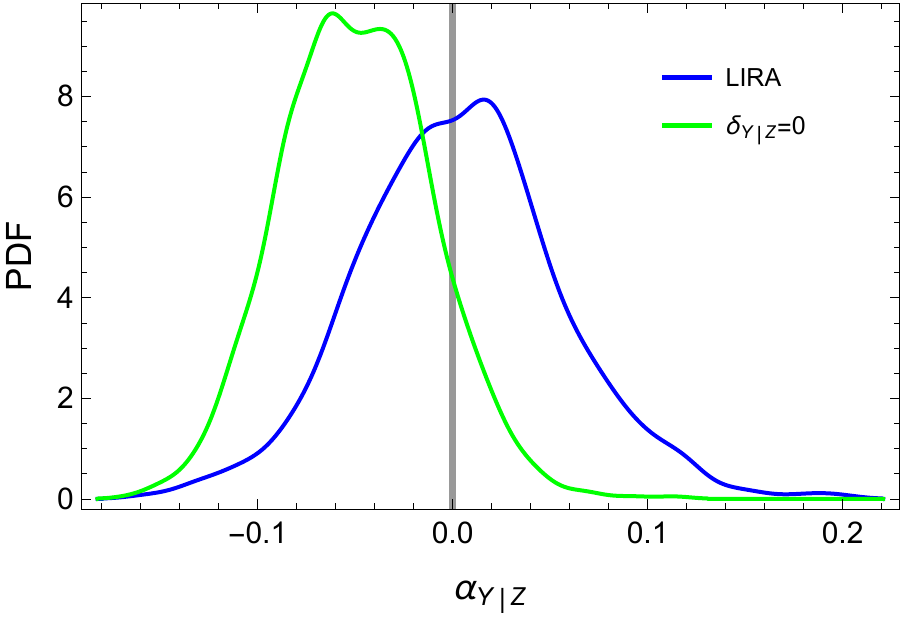}} \\
\resizebox{0.35\textwidth}{!}{\includegraphics{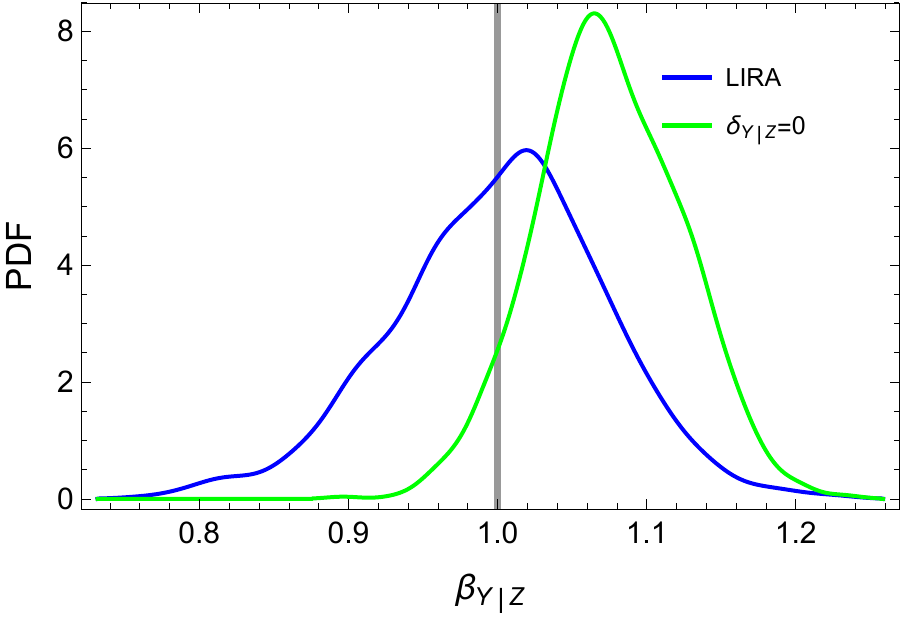}} \\ 
\resizebox{0.35\textwidth}{!}{\includegraphics{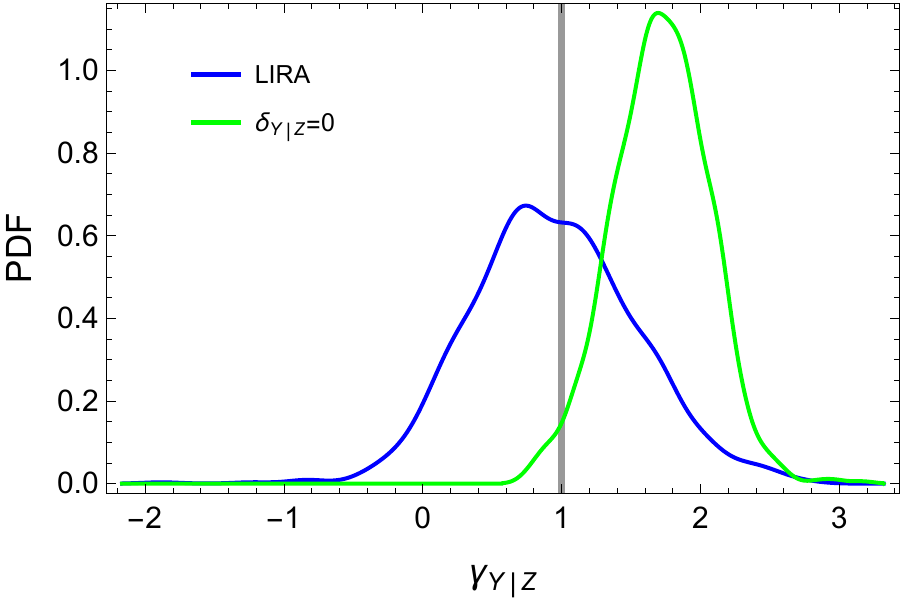}} \\ 
\resizebox{0.35\textwidth}{!}{\includegraphics{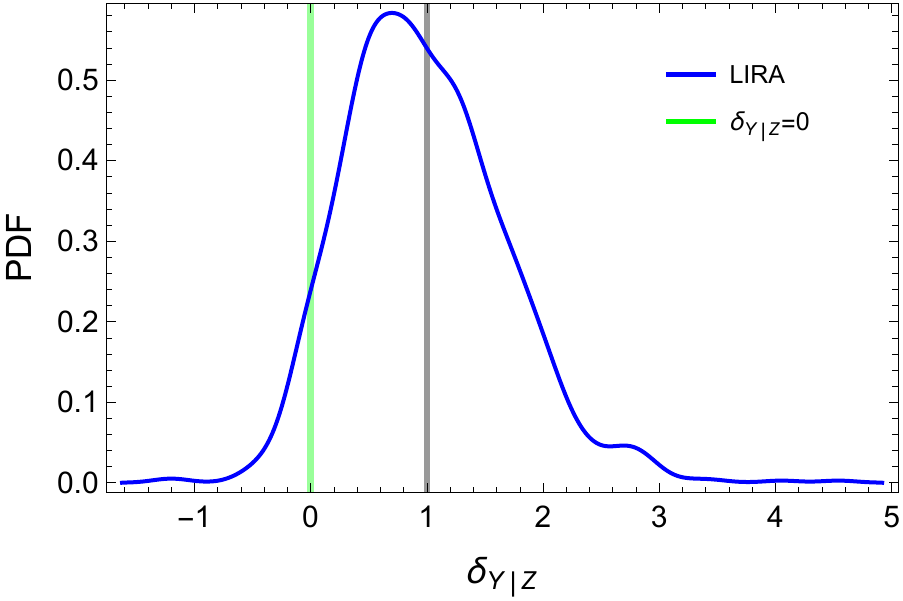}} \\ 
\resizebox{0.35\textwidth}{!}{\includegraphics{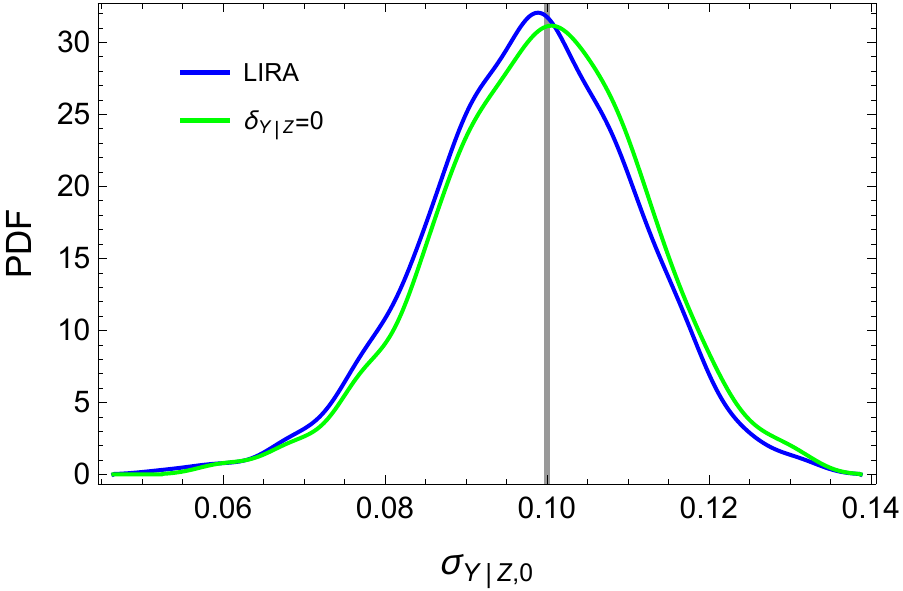}} \\
\end{tabular}
\caption{Distributions of the median parameters obtained from samples with time-evolving slope. The blue lines are the smoothed histograms of the distributions obtained by fitting the simulated data with a time-dependent slope. The green lines plot the results from a biased linear fit with $\delta_{Y|Z}=0$. The vertical gray lines are set at the input parameters. From the top to the bottom panel: the intercept, the slope, the time evolution, the slope tilt, and the intrinsic scatter.}
\label{fig_pdf_deltaYIZ}
\end{figure}

The emergence of some processes at high or low redshift might induce a tilting slope. I simulated a scaling relation with $\delta_{Y|Z}=1$. In this case, the slope changes by $\Delta \beta_{Y|Z}\sim0.25$ from redshift 0 to 1.

Redshifts were drawn from a lognormal distribution such that $\ln z$ has mean $\ln (0.3)$ and standard deviation 0.5. The independent variables were drawn from a time evolving normal distribution with $\gamma_{\mu_Z,F_z}=0.5$  and $\gamma_{\mu_Z,D}=0.5$; the standard deviation is constant. The scale parameters of the scaled inverse $\chi^2$-distributions modeling the uncertainty variances $\delta x^2$ and $\delta y^2$ were set to $0.05^2$ and measurement errors were assumed to be uncorrelated. The remaining parameters were set as in the basic scheme.

Results are summarized in Table~\ref{tab_sim_time} and Fig.~\ref{fig_pdf_deltaYIZ}. A correct modeling of the tilt is crucial to get unbiased parameters. Only the estimate of the intrinsic scatter is not affected.

\section{Conclusions}
\label{sec_conc}

Bayesian linear regression models can involve a large number of parameters. The analysis of the hierarchical models can be performed with Markov Chain Monte Carlo (MCMC) simulations. Since all relations in the model are expressed as conditional probabilities, the posterior can be efficiently explored with a Gibbs sampler \citep{kel07,man15}. 

\textsc{LIRA} joins a number of already available routines for linear regression. Just to name a few of them which were proposed to astronomers first, the Fortran function \textsc{BCES}, the IDL (Interactive Data Language) routine \textsc{LINMIX} and its multivariate extension \textsc{MLINMIX}, the Python package \textsc{astroML}\footnote{\url{http://www.astroml.org/}}\citep{van+al12}, and the R-packages \textsc{LRGS}\footnote{\url{https://cran.r-project.org/web/packages/lrgs/index.html}}, and \textsc{HYPER-FIT}\footnote{\url{https://github.com/asgr/hyper.fit}}.

All of these procedures have their own specifics and strengths that can make them preferable under given circumstances. \textsc{LIRA} is optimized for astronomical studies. It allows the consistent treatment of time-evolution, intrinsic scatter, and selection effects. Redshift has a prominent role in the proposed method. The time dependence of slopes, normalizations, intrinsic scatters, and correlations can be determined. Further complexity is implemented. The Malmquist and the Eddington biases can be addressed. Deviations from linearity and bent relations with knees can be accounted for.

The degree to which selection and methodological biases can affect the study of current and future samples was determined with a series of simulations. Selection effects are an important concern. But they are known problems and to some extent they are known unknowns. We usually know whether they are affecting our sample. Methodological biases can be unknown unknowns. Different parameterizations can give excellent fits to the data with significantly different results. We do not know a priori the right parameterization. The problem is exacerbated by the high degree of degeneracy among involved parameters. The feature of a linear regression model to stay simple and to add complexity if needed is then important.

\section*{Acknowledgements}
I thank S. Ettori for encouragement and stimulating discussions. M.S. acknowledges financial contributions from contracts ASI/INAF n.I/023/12/0 `Attivit\`a relative alla fase B2/C per la missione Euclid', PRIN MIUR 2010-2011 `The dark Universe and the cosmic evolution of baryons: from current surveys to Euclid', and PRIN INAF 2012 `The Universe in the box: multiscale simulations of cosmic structure'. S.E. acknowledges the financial contribution from contracts ASI-INAF I/009/10/0 and PRIN-INAF 2012 `A unique dataset to address the most compelling open questions about X-Ray Galaxy Clusters'.

\footnotesize
%\bibliographystyle{mn2e_fix_Williams}
%\bibliography{/Users/maurosereno/Documents/bozze/all_references_v02}

\setlength{\bibhang}{2.0em}

\normalsize

\appendix

\section{The \textsc{LIRA} R-package}
\label{app_pack}

The package \textsc{LIRA} is publicly available from CRAN\footnote{\url{https://cran.r-project.org/web/packages/lira/index.html}} or GitHub\footnote{\url{https://github.com/msereno/lira}}. It can be installed from within R with the following command

\noindent \texttt{> install.packages("lira",~dependencies=TRUE)}
%\noindent \texttt{> install\_github("msereno/lira")}

\noindent \textsc{LIRA} relies on the \textsc{JAGS} (Just Another Gibbs sampler) library\footnote{JAGS by M. Plummer is publicly available at \url{http://mcmc-jags.sourceforge.net}.}, which must be installed separately, to perform the Gibbs sampling. C++ compilers are also needed.

The package is loaded into the R-session with
 
\noindent \texttt{> library(lira)}

The linear regression analysis is performed through the function \texttt{lira} (hence the name of the package), whose output are Markov chains produced with a Gibbs sampler. Let \texttt{x}, \texttt{y}, \texttt{delta.x}, \texttt{delta.y}, \texttt{covariance.xy}, and \texttt{z} be the vectors storing the values of $\mathbfit{x}$, $\mathbfit{y}$, $\mathbfit{$\delta_x$}$, $\mathbfit{$\delta_y$}$, $\mathbfit{$\delta_{xy}$}$ and $\mathbfit{z}$, respectively.

\begin{itemize}

\item The basic regression in Sec.~\ref{sec_sim_skew} was performed with

\noindent \texttt{> mcmc <- lira(x, y,~delta.x=delta.x, delta.y=delta.y, covariance.xy=covariance.xy)},

\item The Gaussian mixture in Sec.~\ref{sec_sim_skew} was implemented as

\noindent \texttt{> mcmc <- lira(x, y,~delta.x=delta.x, delta.y=delta.y, covariance.xy=covariance.xy, n.mixture=3)}

\item The chains analyzed in \ref{sec_sim_skew_z} to address a time-dependent skewed distribution $p(Z,z)$ were obtained with

\noindent \texttt{> mcmc <- lira(x, y,~delta.x=delta.x, delta.y=delta.y, covariance.xy=covariance.xy, z=z, gamma.sigma.Z.Fz="dt")},

\item The case of the Eddington bias in Sec.~\ref{sec_sim_eddi} was studied with

\noindent \texttt{> mcmc <- lira(x, y,~delta.x=delta.x, delta.y=delta.y, covariance.xy=covariance.xy, sigma.XIZ.0="prec.dgamma")}

\item The regression corrected for Malmquist bias, as in Sec.~\ref{sec_sim_malm}, is performed with

\noindent \texttt{> mcmc <- lira(x, y,~delta.x=delta.x, delta.y=delta.y, covariance.xy=covariance.xy, y.threshold = rep(-0.3, n.data))}

\noindent where  \texttt{n.data} is the length of the vectors. 

\item The broken power-law in Sec.~\ref{sec_sim_brok} was analyzed with

\noindent \texttt{> mcmc <- lira(x, y,~delta.x=delta.x, delta.y=delta.y, covariance.xy=covariance.xy,  Z.knee="dunif(-3.0,3.0)", beta.YIZ.knee ="dt")}

\item The scaling with time dependent intrinsic scatter in Sec.~\ref{sec_sim_scat} was analyzed with

\noindent \texttt{> mcmc <- lira(x, y,~delta.x=delta.x, delta.y=delta.y, z=z, gamma.sigma.XIZ.Fz="dt")}

\item The scaling with time evolving slope in Sec.~\ref{sec_sim_tilt} was analyzed with

\noindent \texttt{> mcmc <- lira(x, y,~delta.x=delta.x, delta.y=delta.y,  z=z,  delta.YIZ="dt")}

\end{itemize}

The \textsc{LIRA} package and further material and examples can also be found at \url{http://pico.bo.astro.it/\textasciitilde sereno/LIRA/}.

\section{Redshift evolution}
\label{app_reds}

In \textsc{LIRA}, the time-evolution of the parameters is factorized in two terms, one depending on $F_z$ and one on the distance $D$. The factor $F_z$ can be either $E_z$ or $(1+z)$. Since the redshift evolution is poorly constrained in present data-sets, and since both the cosmological distances and $F_z$ are increasing function of the redshifts, the estimates of the evolution parameters $\gamma_{F_z}$ and  $\gamma_{D}$ of each scatter/dispersion parameter are highly degenerate. It is usually enough to model just one dependence. The exception is the time evolution of the means in the mixture modeling $p(Z)$.

For limited redshift baselines, the function $E_z$ can be approximated with a power law of $(1+z)$. The value of the exponent used in the approximation depends on the redshift range considered and on the cosmological parameters. In most cases, modeling $F_z$ as either $E_z$ or $(1+z)$ has a minor effect on the estimates of the scaling parameters $\alpha$ and $\beta$ and of the intrinsic scatters.

For similar reasons, the choice of the cosmological distance is usually secondary. The angular diameter and the luminosity distance differ for a factor $(1+z)^2$ which can be approximately englobed in $E_z^{\gamma_{F_z}}$ for limited redshift baselines. Luminosity distances can be preferred for flux limited samples.

\end{document}